\documentclass[prd,
 reprint,
 amsmath,amssymb,
 aps,nofootinbib
]{revtex4-2}
\usepackage{bm}
\PassOptionsToPackage{linktocpage}{hyperref}
\usepackage[hyperindex,breaklinks,hidelinks]{hyperref}
\usepackage{enumitem}
\usepackage{slashed}
\usepackage[dvipsnames]{xcolor}

\usepackage{subcaption}

\usepackage{microtype}

\usepackage[compat=1.1.0]{tikz-feynman}

\renewcommand{\vec}[1]{\ensuremath{\boldsymbol{#1}}}

\usepackage{tikz}
\tikzset{every picture/.style={line width=0.75pt}} 
\usetikzlibrary{shapes.geometric}
\usepackage{orcidlink}
\usepackage{comment}
\usepackage{array}
\usepackage{mathtools}

\usepackage{etoolbox}

\renewcommand{\vec}[1]{\ensuremath{\boldsymbol{#1}}}

\newcommand{\bra}[1]{\ensuremath{\left< #1\,\right|}}
\newcommand{\ket}[1]{\ensuremath{\left|\, #1\right>}}

\def\d{\text{d}}

\begin{document}

\title{The Role of Microstate Degeneracy in Phase Transitions: \\
\textit{Gravitational Waves from Bubble Entanglement}}

\author{Gia Dvali} 
\email{gdvali@mpp.mpg.de} 
\author{and Lucy Komisel\,\orcidlink{0009-0008-1578-588X}} 
\email{lucy.komisel@mpp.mpg.de}
\affiliation{Arnold Sommerfeld Center, Ludwig-Maximilians-Universit\"at, Theresienstra{\ss}e 37, 80333 M\"unchen, Germany}
\affiliation{Max-Planck-Institut f\"ur Physik, Boltzmannstr. 8, 85748 Garching, Germany}

\date{\today}

\begin{abstract} 
  
Vacuum bubbles, formed in first order phase transitions, have important implications for cosmology. 
In particular, they source gravitational waves. 
Usually, it is assumed that, once bubbles are materialized, their state, further evolution and mergers are well-described classically. 
This paper will show that this intuition breaks down for bubbles which possess high microstate degeneracy. This is generic when the phase transition breaks spontaneously a symmetry.  
First, the degeneracy enhances the transition rate. 
Furthermore, the internal quantum state of the bubbles profoundly affects the classical dynamics of their mergers. 
A bubble, no matter how macroscopic,  is born in a maximally entangled quantum state. 
This state can be viewed as a symmetric superposition of many different would-be classical bubbles. 
The inner entanglement is largely maintained up until their mergers. 
The resulting true quantum dynamics of the merger is macroscopically different from any type of classical mergers. 
These differences are imprinted as macroscopic features in the resulting classical gravitational waves. 
In this way, the inner microstate entanglement of merging bubbles provides a qualitatively new source of gravitational waves. 
This phenomenon is quantified and compared with the \textit{swift memory burden effect} in black hole mergers.   

\end{abstract}
 
\maketitle

\section{Introduction}  
Large inhomogeneities play an important role in cosmology and astrophysics. 
In particular, they can serve as sources of gravitational waves.  
Examples are provided by the coalescence of vacuum bubbles nucleated in first order phase transitions, collapsing topological defects or merging black holes. 
  
Usually, it is assumed that these sources are well-described classically. 
This intuition however breaks down for sources that posses a large microstate degeneracy. 

In this paper, we show that this is the case for vacuum bubbles created in a first order phase transition with symmetry breaking. 
Due to the degeneracy of the bubble states, no matter how macroscopic, the bubble 
is materialized in a highly quantum state of a maximally entangled superposition of would-be classical bubbles. 
The subsequent expansion and the bubble merger is not described by a classical evolution but by the entangled quantum superposition of a large number of would-be classical trajectories.  
This provides a qualitatively new source of classical gravitational waves, fundamentally different  from any type of a classical source. 
   
The macroscopic influence of the internal quantum degeneracy on the classical dynamics of the bubble merger shares similarities with the so-called \textit{swift memory burden effect}~\cite{Dvali:2025sog} in mergers of black holes, which are  well-known  objects with a high microstate degeneracy. 
The essence of the phenomenon is that the internal quantum state of a black hole strongly affects its classical dynamics when the black hole is perturbed. 
Due to its large macrostate degeneracy, classically identical back holes can be in very different internal quantum states due to differences in the information load that they carry. 
Upon perturbation, this internal difference leads to vastly different classical evolutions. 
In particular, the higher information load ``stiffens" the host, making a black hole more resistant against perturbations. 

That is, a black hole, on top of standard quantum numbers such as the mass, the angular momentum and the electric/magnetic charge, possesses a further macroscopic quantum characteristic: the \textit{memory burden parameter}, $\mu$. 
This parameter measures the usage of the black hole's memory space.  
In particular, it tells us about the portion of the memory storage occupied by the information load. 
    
When the black hole is in its ground state, this feature is invisible, since black holes of the same mass can carry vastly different information loads. 
However, the memory load swiftly affects the classical dynamics upon a perturbation.  
In particular, this phenomenon  takes place in black hole mergers and influences their gravitational wave spectrum.  
An analogous effect was shown to be exhibited in mergers of solitons of high microstate degeneracy. 

Originally~ \cite{Dvali:2018xpy, Dvali:2018ytn, Dvali:2020wft}, the \textit{memory burden effect} was applied to the quantum evolution of systems and, in particular, was shown to stabilize them against decay.  
For example, the back-reaction from the information load is expected to slow down the decay of a black hole via the Hawking evaporation. 
For a large black hole, entrance into the memory burden phase via evaporation happens gradually on a macroscopically long time scale.  
In contrast, swift memory burden sharply influences the macroscopic dynamics of the black hole on a time-scale of the classical perturbation. 
Correspondingly, it is potentially easier to observationally detect the swift memory burden effect in black hole mergers via its imprints in gravitational radiation. 
  
In this sense, the swift memory burden effect is an example of  \textit{macro-quantumness} \cite{macroQ}, a concept originally introduced in a quantum description of black holes \cite{Dvali:2011aa}.
However, as discovered later \cite{S1,S2,Saturon2020,Dvali:2021jto, S3,S4,S5,S6, Dvali:2024hsb, Contri:2025eod}, this feature is not an exclusive property of black holes but extends to a much wider class of quantum field theoretic (QFT) objects with large microstate degeneracy. 
This is because such object operate via the mechanism of \textit{assisted gaplessness}~\cite{Dvali:2017nis, Dvali:2017ktv, Dvali:2018vvx, Dvali:2018xoc, Dvali:2018tqi},\cite{Dvali:2018xpy, Dvali:2018ytn, Dvali:2020wft}: in their interior, they create a local environment in which a macroscopically large number $N$ of degrees of freedom becomes gapless.  
These are called \textit{memory modes} since the diversity of their excitation patters (\textit{memory patterns}) measures the information storage capacity of the system. 
Due to the gaplessness, the patterns are degenerate in energy creating a basis of the \textit{memory space}.  
  
The diversity of the memory modes $N$, and thus the corresponding microstate entropy, $S$, is constrained by a universal QFT upper bound, derived in~\cite{S1, S2, Saturon2020}.
Objects close to saturating this bound are called \textit{saturons} \cite{Saturon2020}.   
Black holes are the most prominent representatives of this universality class, which however is much wider.  
Previous studies show that members of the saturon family share key characteristics with black holes. 
 
In the present paper,  we study vacuum bubbles formed in first order phase transitions. 
As we shall see, vacuum bubbles of high microstate degeneracy exhibit some similarities with the swift memory burden effect of black holes in the sense that the internal quantum state of the bubbles macroscopically affects their mergers and the resulting gravitational waves. 

Some key ingredients of the framework have been prepared by previous studies. 
Vacuum bubbles exist in theories in which the minima of a scalar potential are separated by a potential energy barrier. 
If two neighbouring minima have the same energy, the bubbles are non-expanding. 
In the opposite case, a bubble of the ``true" vacuum,  larger than a certain critical radius, expands unbounded.  

Non-expanding vacuum bubbles of high microstate degeneracy have been studied in \cite{Saturon2020, Dvali:2021tez, Dvali:2024hsb}. 
These studies have shown many similarities between saturated bubbles and black holes. 

The present paper shall focus specifically on high degeneracy vacuum bubbles produced in phase transitions via thermal or quantum tunneling and shall analyze their subsequent mergers.  
It will be shown that this leads to qualitatively new phenomena. 
Namely, the microstate degeneracy  greatly influences all three stages of the phase transition:  
{\it 1)} the bubble creation rate, 
{\it 2)} the quantum state of the materialized bubbles and 
{\it 3)} the dynamics of their mergers.

In order to be more precise, let us consider a first order phase transition, which proceeds via the nucleation of bubbles of the lower energy vacuum.
In the standard treatment, the microstate degeneracy of the nucleated bubble is not considered. 
In such a case, even though the bubble nucleation process is quantum, it is assumed that the bubbles quickly classicalize.  
In such a treatment the internal quantum entanglement of the materialized bubble is neglected.

Now, taking into account the bubble microstate degeneracy alters the story in the following way. First, as already discussed in \cite{SaturonDM}, it enhances the rate of bubble nucleation. 
In the extreme case, the microstate entropy of the bubble can compensate the exponential suppression.  

As shown in \cite{S2}, in the quantum tunneling process, there exists a generic correlation between the saturation of the QFT entropy bounds by the materialized Lorentzian object and its corresponding Euclidean instanton. 
In particular, the saturation was shown to be exhibited by gauge instantons when the theory approaches the confining regime.  

In the present paper we observe a similar correlation in the decay of the false vacuum.
Namely, the QFT bounds on the microstate entropy are saturated simultaneously (i.e., for the same values of the parameters) by a critical bubble and the corresponding Euclidean bounce. 
 
The new key feature is that the degenerate vacuum bubble materializes in a superposition of an exponentially large number $\sim {\rm e}^{S}$ of microstates.
In this superposition, the internal degrees of freedom of the bubble are maximally entangled. 
This state cannot be described classically. 
Due to very high initial entanglement, by the time the two bubbles overlap, they are fully quantum objects.
The generation of gravitational waves from this highly entangled matter is macroscopically different from ordinary classical mergers. 

In the previous analyses of gravitational waves sourced by colliding bubbles \cite{Kosowsky:1991ua, *Kosowsky:1992rz, FirstOrder1}, collapsing defects \cite{Vilenkin:1981bx, Vachaspati:1984yi, Martin:1996ea, Martin:1996cp, Gleiser:1998na}, or various forms of turbulence \cite{Turbulence1, Turbulence2, Turbulence3}, the effect of strong internal entanglement has not been considered. 
 
Objects with internal microstate degeneracy exhibit a very different behaviour due the feature of macro-quantumness.  
This applies equally to merging black holes as well as to degenerate vacuum bubbles formed in phase transitions.
The internal microstate degeneracy allows them to be simultaneously macroscopic and quantum.
In what follows, we shall study the dynamics and implications of this phenomenon. 
      
\section{QFT Bound on Microstate Entropy} 

Before discussing the role of the microstate degeneracy in phase transitions, we briefly review the QFT constraints on this degeneracy obtained in~\cite{S1, S2, Saturon2020}.
The level of degeneracy is quantified by the microstate entropy, defined as the logarithm of the number of degenerate microstates,  $S \equiv \ln(n_{st})$.     
As shown in~\cite{S1, S2, Saturon2020}, QFT imposes a certain universal upper bound on  the microstate entropy of any object, which we shall now review.

Let us consider a  QFT  formulated in terms of certain degrees of freedom $\phi_j$ (labeled by $j=1,2, ...$) interacting via a quantum coupling $\alpha$. 
The coupling is of course ``running" in the sense that it depends on the characteristic scale of a process. 

The necessary condition for the QFT to be valid is that the coupling is weak, $\alpha \ll 1$, which will be assumed. 
However, this condition alone is not sufficient, since the degrees of freedom can be invalidated due to certain collective phenomena, such as multi-particle processes. 
  
We shall assume that the above degrees of freedom form a localized macroscopic bound state, with a localization radius $R$. 
For example, this could be a soliton. 
For definiteness, it will be assumed that the object is approximately spherical.  
 
Under these conditions, the validity of the QFT description in terms of the $\phi$ degrees of freedom puts a strict upper bound on the microstate entropy of the object. 
This bound can be written in two equivalent ways~\cite{S1, S2, Saturon2020}.

The first form is:    
\begin{equation}  \label{Alpha} 
    S \, \leqslant \,  S_{\rm max} \equiv \frac{1}{\alpha} \,.  
\end{equation} 
In the above expression, $\alpha$ has to be understood as the scale-dependent (running) coupling evaluated at the scale (momentum-transfer) $1/R$.

Alternatively, the bound can be written in terms of the scale $f$ of spontaneous breaking of Poincar\'e symmetry.
Notice that any localized entity breaks a part of the Poincar\'e symmetry  spontaneously. 
At the very least, this applies to translations and the Lorentz boosts.
In each case, the scale $f$ is uniquely defined. 
Of course, the spontaneous breaking of Poincar\'e symmetry is linked to the existence of gapless Nambu-Goldstone modes. 
The scale $f$ determines the coupling of these Goldstones as $G=f^{-2}$.
The dimensionless coupling of the canonically-normalized Poincar\'e Goldstone is $\alpha_{\rm gold} = 1/(\pi R^2 f^2)$.
    
In terms of the Poincar\'e-breaking scale, the bound on the microstate entropy can be written as,  
\begin{equation}  \label{Area} 
    S \, \leqslant \, \frac{Area}{4G} 
    \, = \, \pi R^2 f^2   \,,   
\end{equation} 
where, $Area = 4\pi R^2$ is the surface area of the object. 
Unsurprisingly, the equation \eqref{Area}, written in terms of the dimensionless coupling of the Poincar\'e Goldstone, coincides with \eqref{Alpha}.

The expression \eqref{Area} is strikingly similar to the Bekenstein-Hawking entropy of a black hole~\cite{Bekenstein:1973ur}. 
In fact, this similarity is not a coincidence but reveals a deep physical connection.  

This connection can be captured by noticing that in the case of a black hole the scale of Poincar\'e breaking is the Planck scale, $f = M_P$ (see, e.g.,~\cite{Dvali:2025sog}). 
This is because the deviation from the asymptotic Minkowski metric (measured by the expectation value of the canonically-normalized graviton field) becomes of order $M_P$ near a black hole horizon.
This happens irrespective of the mass of a black hole.
At the same time, such a high value of the gravitational field is never reached in the vicinity of any object that is spread outside of its gravitational radius. 
In this sense a black hole can be defined as the object that breaks Poincar\'e symmetry at the Planck scale. 

Correspondingly, the Goldstone coupling for a black hole is given by Newton's constant $G_N =1/M_P^2$. 
This suggests that the black hole entropy represents a particular case of saturation of the more general entropy bound \eqref{Area}. 
This bound is not specific to gravity, but it is in fact universal to objects in arbitrary theories. 
    
The bounds \eqref{Alpha} and \eqref{Area} were derived by correlating the microstate degeneracy of a bound state to the validity of the QFT description~\cite{S1, S2, Saturon2020}. 
In particular, the saturation of these bounds takes place at the verge of validity of the loop expansion in powers of $\alpha$. 
At the same time, it is linked to the saturation of unitarity by multi-particle scattering amplitudes. 
For a detailed discussion, the reader is referred to the above original articles.  

It is important to point out that to constrain the microscopic degeneracy in phase transitions, we must apply the bounds \eqref{Alpha} and/or \eqref{Area} rather than the more widely known Bekenstein bound on entropy: 
\begin{equation}  \label{Bek} 
    S = 2\pi ER,  
\end{equation} 
where $E$ is the energy of the object. 
The reason is that the Bekenstein bound is ill-defined both for vacuum bubbles, which can have zero energy, as well as for instantons~\cite{S2}, which are Euclidean entities with no notion of energy. 
In contrast, the bounds \eqref{Area} and \eqref{Alpha} are fully applicable to vacuum tunneling.  
Indeed, both the running coupling $\alpha$ and the scale of Poincar\'e-breaking $f$ are well defined quantities for Euclidean instantons as well as for Lorentzian vacuum bubbles. 
Thereby, 
to constrain the degeneracy of such objects, one has to use the expressions \eqref{Alpha} and \eqref{Area}.  
These expressions will suffice to capture the effects of the microstate degeneracy in phase transitions. 

\section{Phase Transition for High Microstate Degeneracy} 
 
\subsection{The Setup}
  
It is now time to discuss the peculiarities of phase transitions in case of a high microstate degeneracy. 
For this, we consider an explicit example of a theory supporting a vacuum bubble with the microstate entropy arbitrarily close to saturation.  

These bubbles were originally introduced in \cite{Saturon2020}, and their production at high temperature was studied in \cite{SaturonDM}.    
For the present discussion, a slightly altered model compared to the previous papers will be used.

The system involves a real scalar field $\phi_j, ~ j=1,2,...,N$ transforming in the $N$-dimensional representation of an $O(N)$ global symmetry.  

The high-temperature potential is chosen in the form, 
\begin{align}  \label{V1}
    V(\phi, T)\,  = &  - \frac{M(T)^2}{2} \phi^2  +  \nonumber \\ 
    + & 
    \frac{\alpha}{2f^2} \phi^2  \left ( \phi^2 - f^2 \right )^2\,.     
 \end{align} 
where  $\phi^2 \equiv \phi_j\phi_j$, the parameter $f$ is the scale of symmetry breaking, and $\alpha$ is a dimensionless coupling constant. 
The non-renormalizability is not an issue here, as it is always possible to stay within regimes where the EFT approach is valid. 
Furthermore, the above theory can easily be completed into a renormalizable model of \cite{SaturonDM} by integrating-in an additional singlet field. 
Alternatively, all our conclusions can be derived by using the two-index symmetric representation of $O(N)$ or the adjoint of $SU(N)$, instead of the fundamental as was originally done in~\cite{Saturon2020}. 
In this case, the potential describing the vacuum bubbles with spontaneous symmetry breaking is renormalizable. 
We present this construction in the Appendix.  
  
The region of interest is  $\alpha \ll 1$, ensuring that the low energy EFT is in the weak-coupling regime.   
However, the validity of the EFT description in terms of $\phi$ puts the following bound on the parameters of the theory, 
\begin{equation} \label{Ubound}
  \alpha  \lesssim \frac{1}{N} \,.  
\end{equation} 
This bound can be understood by noticing that the expansion parameter coming from the quartic coupling, $\alpha (\phi^2)^2$ is $\alpha N$, since each loop factor contains a sum over all $N$ flavors of $\phi_j$. 
 
The largest expansion parameter, coming from the sixth-order vertex  $\frac{\alpha}{f^2} (\phi^2)^3$, is $\alpha \frac{\Lambda^2}{f^2} N^2$, where $\Lambda$ is the EFT cutoff.
This cutoff is bounded from below by the mass of the $\phi$-particles, $m = \sqrt{\alpha} f$, giving, 
\begin{equation} \label{Lcutoff}
    \Lambda^2  > \alpha f^2 \,. 
\end{equation}
From here it follows that the  expansion parameter is bounded from below by $\alpha^2N^2$. 
This quantity must be smaller than 1 for the EFT to be valid. 
This imposes the bound given in Eq.\eqref{Ubound}. 
An analogous bound is imposed by investigating scattering amplitudes involving $\sim N$ quanta of $\phi$'s with momentum transfer per particle $\sim m$ \cite{Saturon2020}. 

\begin{figure}
    \begin{subfigure}{\linewidth}
        \begin{tikzpicture}
        \def\r{1}
        \begin{feynman}
            \vertex (a) at (-2*\r,0){$\phi_j$};
            \vertex (b) at (2*\r,0){$\phi_j$};

            \diagram{
                (a) -- (b)};
        \end{feynman}
        \end{tikzpicture}
        \caption{}
        \label{fig:withoutbutterfly}
    \end{subfigure}
    \begin{subfigure}{\linewidth}
        \begin{tikzpicture}
        \def\r{1}
        \begin{feynman}
            \vertex (a) at (-2*\r,0){$\phi_j$};
            \vertex[dot] (p) at (0,0){};
            \vertex (pl) at (0,0.5*\r){$\propto \alpha$};
            \vertex (lu) at (0,\r);
            \vertex (lul) at (0,1.5*\r){$\propto N$};
            \vertex (ld) at (0,-\r);
            \vertex (ldl) at (0,-1.5*\r){$\propto N$};
            \vertex (b) at (2*\r,0){$\phi_j$};

            \diagram{
                (a) -- (p) -- (b),
                (p) --[half left] (lu) --[half left] (p),
                (p) --[half left] (ld) --[half left] (p)};
        \end{feynman}
        \end{tikzpicture}
        \caption{}
        \label{fig:withbutterfly}
    \end{subfigure}
    \caption{The diagram describing the sensitivity w.r.t. $N$ in form of a  "butterfly" insertion (Fig. 3b) to an arbitrary line in Feynman diagram (Fig. 3a). 
    Each such insertion gives a relative factor $\propto \alpha N^2\frac{\Lambda^2}{f^2}$, since each butterfly "wing" contributes a factor $N$. 
    The validity of the loop-expansion in EFT below the cutoff scale $\Lambda$ demands this factor to be less than one. 
    This diagram was created using TikZ-Feynman \cite{ELLIS2017103}}
    \label{fig:butterfly}
\end{figure}
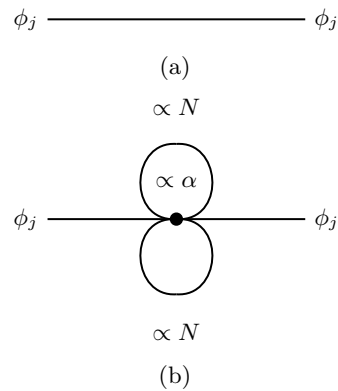

Notice however, that the expansion parameter $\frac{\Lambda^2}{f^2}\alpha N^2$, which comes from attaching a two-loop ``butterfly" extension via a $ \frac{\alpha}{f^2} (\phi_j^2)^3$ vertex to any line in a Feynman diagram, has to be smaller than one, in order for the  EFT loop-expansion to remain valid.  This is illustrated in Fig.\ref {fig:withoutbutterfly},\ref{fig:withbutterfly}.

Therefore, the regime of validity for the EFT is,  
\begin{align}\label{cutofflimits}
    m <\Lambda <\frac{m }{\alpha N}\,.
\end{align}
This expression shows that in the saturation regime, where $N\alpha \sim 1$, the theory starts entering a new regime. 
    
Saturon bubbles exist, provided the theory saturates the bound \eqref{Ubound}. 
However, in order to fully appreciate the importance of the microstate degeneracy for phase transition and gravitational waves, the saturation of the degeneracy is not required. 
 
Instead, we shall allow $N$ to vary between the extreme cases of $N \sim 1/\alpha$ and $N=1$. 
In the latter case, the symmetry group of the theory shrinks to a discrete $Z_2$-symmetry acting on a real scalar field, $\phi \rightarrow -\phi$ and the microstate degeneracy of the bubble is minimal.  
 
Next, the origin of the thermal mass term $M^2(T)$ in \eqref{V1} will be elaborated on. It is assumed, that this correction to the mass is coming  from the interaction of $\phi$  with some species that are in thermal  equilibrium  at temperature $T$.
It is sufficient to discuss a single $O(N)$ singlet scalar field $\chi$, which interacts with $\phi$ via the following coupling in the potential, 
\begin{align}  \label{PhiChi}
    - \beta \chi^2 \phi^2 ,
\end{align} 
where $\beta > 0$ is a coupling constant.  
Similarly to \eqref{Ubound}, the validity of the EFT description imposes the following bound on $\beta$,
\begin{align} \label{BetaB}
   \beta \lesssim \frac{1}{N}, 
\end{align}
which shall be assumed to be satisfied. 
    
For definiteness, we shall assume that the $\phi_j$ are not in thermal equilibrium.  
Of course, in the absence of external factors, for any finite value of the cross-coupling $\beta$, the two subsystems will sooner or later thermalize and equilibrate.
In this light, the statement that $\phi$ is not in thermal equilibrium means that the time-scale of interest is much shorter than the equilibration time, which scales as  the inverse scattering rate. 
For example, in the cosmological context this implies that the $\chi - \phi$ and $\phi - \phi$ scattering rates are less than the Hubble expansion rate.  
  
Even so, the coupling of $\phi$-s with the thermal bath of $\chi$-s generates an effective mass term for $\phi$, 
\begin{align}  \label{MTChi}
      M^2(T) \sim \beta T^2 . 
\end{align} 
This is the case even if the occupation number of $\phi$-quanta is zero.  
  
In order to make the role of different parameter regimes transparent, it is useful to consider the following extreme limits.  

\subsubsection{Decoupling Regime} 
        
The ``decoupling limit" is the following double scaling limit: 
\begin{align}  \label{DSlimit}
    T \rightarrow \infty\,,~  \beta \rightarrow 0\,, ~  
    (\beta T^2) = {\rm finite} \,, ~   N \beta  \rightarrow 0\, \,. 
\end{align} 
In this limit, the scattering rate of $\chi$-quanta 
into $\phi$-s vanishes, 
\begin{align}  \label{CPrate}
     \Gamma_{\chi \rightarrow \phi}  \sim \beta^2TN  
  \rightarrow 0 \,, 
\end{align} 
meaning that the two sectors decouple at the microscopic quantum level
\footnote{Of course, in a cosmological context, it is simultaneously necessary to take $M_p \rightarrow \infty$, in order for gravity not to play an essential role in the transition.}. Furthermore, in this limit, the time scale of thermaliaztion of $\phi$ goes towards infinity, due to the two fields decoupling.
However, the collective thermal effects remain.  
In particular, the thermal mass of $\phi$, given in \eqref{MTChi}, remains non-zero. 
 
The decoupling limit leaves $N$ and $\alpha$ not fully specified, which can be in various regimes. In particular, in addition to \eqref{DSlimit} it is possible to take the 't Hooft-like large-$N$ limit, 
\begin{align}  \label{TNlimit}
    N \rightarrow \infty\,,~  \alpha \rightarrow 0\,, ~  
    N\alpha  = {\rm finite}\,. 
\end{align}

\subsubsection{Common Scaling} 
   
Another regime is that of a common scaling of the two couplings, $\alpha \, \sim  \, \beta$. 
Of course, it is assumed that both coupling are weak and below their unitarity bounds \eqref{Ubound} and \eqref{BetaB}. This regime thereby allows to take a 't Hooft like limit for both couplings, 
\begin{align}  \label{CommonL}
   & T \rightarrow \infty\,,~  \beta \rightarrow 0\,, ~  \alpha \rightarrow 0\,,  \\ \nonumber 
   & (\beta T^2) = {\rm finite} \,, ~  
  N\alpha \sim  N \beta  = {\rm finite}  \,. 
\end{align} 
In this limit the $\chi$-to-$\phi$ scattering rate \eqref{CPrate} vanishes, but the two sectors do not fully decouple, since the effects controlled by collective couplings are non-zero.
This includes, among others, loop corrections from $\phi$-loops in the $\chi$ sector. 
  
As usual, the usefulness of such regimes is that certain statements can be made exact.
These statements can then be extended, in a controlled manner, into domains with finite values of the coupling, using perturbation theory.  

One needs to keep in mind however, that in both scaling regimes, there is some ambiguity left, regarding the scale of symmetry breaking $f$. 
There are two distinct possibilities to consider in the limit \eqref{TNlimit}.

The first possibility is that $f$ stays fixed, while correspondingly $m=\alpha f^2 \rightarrow 0$ as $\alpha \rightarrow 0$.
The second possibility is that $m = \alpha f^2$ stays fixed, with $f^2 \sim N$\footnote{Due to this, the vacuum expectation value $\langle \phi\rangle \sim f$ also scales with $N$ and therefore diverges in this limit.}.

At first glance, both limits seem equally valid, however if $f$ stays fixed, it is straightforward to see from Eq. \eqref{cutofflimits} that the theory actually breaks down for any value of the collective coupling $N\alpha$, as both (the lower and the upper) bounds approach zero as $m\rightarrow 0$. 
Therefore, it is necessary that $m$ stays fixed in both scaling regimes.
\subsection{Vacuum Bubbles}  
 
In the outlined setup, it is assumed that the system starts out in the symmetric state $\phi =0$ with a subsequent bubble of the broken symmetry phase $\phi \neq 0$ forming.
This will be studied in detail when we discuss tunneling. 
However, some relevant bubble configurations will be studied first.  
 
A spherically-symmetric bubble of an $O(N-1)$-invariant vacuum, embedded in the $O(N)$-invariant one, is described by the following classical configuration, 
\begin{align} \label{bubbleconfig}  
    \phi_j(x) = \rho(r,t) e_j, 
\end{align}   
where $e_i$ is a unit $O(N)$ vector,  $e_ie_i = 1$.  
This configuration solves the following equation of motion, 
\begin{align} \label{eqM}
    \Box \phi_j +  \frac{\partial V}{\partial \phi_j} = 0\, .   
\end{align}  
We consider an ansatz in which no motion in the internal $O(N)$ space takes place. 
That is, the entire time evolution is encoded in the radial profile function $\rho(r,t)$. 
For any given moment of time $t$, the function $\rho(r)$ interpolates from $\rho =0$ at $r=\infty$ to a nonzero value in the bubble interior.
  
For $M(T)^2 = 0$, there does not exist a static bubble solution of the above type for any finite radius $R$.
However, the bubble wall configuration for an infinite radius bubble (a domain wall) can be solved exactly and has the form,  
\cite{Dvali:2022rgx},   
\begin{align}  \label{wall}
    \rho(x) = f \frac{1}{\sqrt{1+ {\rm e}^{-2m x}}}
\end{align}     
where $x$ is the coordinate perpendicular to the wall  located in the $x=0$ plane. 
As previously defined, $m=\sqrt{\alpha} f$ is the mass of the quanta in the unbroken symmetry vacuum.   
  
The tension of the domain wall is given by, 
\begin{align} \label{sigma}
    \sigma  \, = \, \frac{1}{2}\sqrt{\alpha} f^3 \, ,
\end{align}  
whereas its thickness is set by the Compton wavelength of the massive mode, $\delta \sim  m^{-1}$. 
 
The expression \eqref{sigma} for the bubble wall tension is an excellent approximation for finite radius bubbles in the thin-wall regime,  $R \,  \gg \, \delta$.
Furthermore, order of magnitude wise it holds true also for thick-wall bubbles   $R \, \sim \, \delta$.

The  vacuum $\phi \neq 0$ can have additional topological structure, which influences the nature of bubbles. 
The case, $N=1$, which corresponds to $Z_2$-symmetry, $\phi \rightarrow -\phi$, is special in the sense that it gives disconnected vacua with $\phi \neq 0$.  
For $M(T)=0$, the potential possesses the three degenerate  minima, $\phi =0$ and  $\phi=\pm f$. Due to this,  for $M(T)=0$ there exist four types of stable planar domain walls: 
the walls  
\begin{align}  \label{wall0+}
    \phi_{0,\pm}(x) = \pm f \frac{1}{\sqrt{1+ {\rm e}^{-2m x}}} \,,
\end{align}   
interpolating from $\phi_{0,\pm}(-\infty) =0$ at $x = -\infty$ to 
$\phi_{0,\pm}(+\infty) = \pm f$ at  $x = +\infty$, 
and the opposite case
\begin{align}  \label{wall+0}
    \phi_{\pm,0}(x) = \pm f \frac{1}{\sqrt{1+ {\rm e}^{2m x}}} \,,
\end{align}     
with asymptotics $\phi_{\pm,0}(-\infty) = \pm f$ and $\phi_{\pm,0}(+\infty) = 0$. 
  
One can create a set of parallel walls of various sequences.  For example, we can denote by 
$\phi_{\mp,0,\pm} $ a sequence of $\phi_{\mp,0}$ and $\phi_{0,\pm}$ walls located, say, in planes at $x=0$ and $x=L$ respectively.
In this configuration the field $\phi$ interpolates from  $\phi(-\infty) = \mp f$ to $\phi(+\infty) = \pm f$ with an intermediate ``stop" at $\phi =0$ for $0 < x <L$.   
Simple energetic arguments indicate that for $M(T) = 0$, the above two walls repel each other.  
However, the repulsive force is exponentially suppressed for distances $L \gg m^{-1}$. 
   
The story changes for $M(T) \neq 0$.  
The minima $\phi \neq 0$ become energetically more favourable compared to $\phi = 0$.   
The local minimum at $\phi = 0$ exists as long as $M^2(T) < \alpha f^2$.  
Correspondingly, for $M(T) \neq 0$, the walls $\phi_{\mp,0}$ and $\phi_{0,\pm}$ are pushed towards each other due to the pressure difference and form a single wall  $\phi_{\mp,0,\pm} $. 
For  $M^2(T) \ll \alpha f^2$,  this wall can be thought of as being a composite of the two ``elementary" walls $\phi_{\mp,0}$  and  $\phi_{0,\pm}$. 
  
For $N>1$, the $\phi \neq 0$, vacua are not disconnected, but can support topological defects in the form of global strings ($N=2$), global monopoles ($N=3$) or textures ($N=3$).  
     
We shall now investigate the three different types of bubbles of $\phi \neq 0$ vacuum that exist for $M^2(T) \neq 0$. 
 
\subsubsection{Extremal Static Bubble} 
 
The first case is an extremal static bubble configuration, which represents a solution of the radial equation,  
\begin{align} \label{eqST}
    \mathrm{d}_r^2 \phi_j(r) + \frac{2}{r}\mathrm{d}_r \phi_j(r) - \frac{\partial V}{\partial \phi_j} = 0 \,.   
\end{align} 
It is described by a time-independent version of the  spherically symmetric bubble \eqref{bubbleconfig} of the type, 
\begin{align} \label{bubblestatic}  
    \phi_j(x) = \rho(r) e_j \,.  
\end{align}     
The energy of the bubble is given by the integral, 
\begin{align}  \label{energyB} 
    E  = 4\pi \int_0^{\infty} r^2\mathrm{d} r \left ( \frac{1}{2} (\mathrm{d}_r \phi_j)^2 +  V(\phi_j) \right ) . 
\end{align}  
 
The static bubble is in the thin-wall regime when the energy density difference between the vacua $\epsilon \simeq \frac{1}{2} M^2f^2$ is significantly smaller than the height of the potential energy barrier separating them, $\sim \alpha f^4$. 
This condition can be written as,
\begin{align}  \label{TWall}    
    M^2(T) \ll m^2 , \
   {\rm or \ equivalently:\ }   
    T^2 \gg  \frac{\alpha}{\beta} f^2.
\end{align}  
 
In this regime, the radius and the energy of the static bubble can be found by extremizing the following expression w.r.t. the bubble radius $R$ \cite{Coleman:1977py}, 
\begin{align}  \label{energy(R)} 
    E(R) \, = \, 4\pi \left ( - \frac{1}{3} R^3 \epsilon  + \sigma R^2
    \right ).  
\end{align}  
This gives 
\begin{align}  \label{energy(R)1} 
    R_0 = 2\frac{\sigma}{\epsilon} \sim \frac{m}{M^2}, ~~ 
    E_b = \frac{16\pi}{3} \frac{\sigma^3}{\epsilon^2}  
    \sim \frac{m}{\alpha} \frac{m^4}{M^4} \,. 
\end{align}  
Notice that for a thin-wall bubble, the volume energy constitutes $\frac{2}{3}$ of the wall tension energy. 
In general, the two energies are of the same order even for a thick-wall bubble.

The static bubble is not stable with respect to small perturbations in its radius. 
Classically, for $R > R_0$ the bubble expands to infinite size, whereas for $R < R_0$ it collapses and decays \cite{Berezin:1982ur}. 
This leads to the other two types of bubbles. 
 
\subsubsection{Zero-Energy Expanding Bubble} 
 
As mentioned, for $R> R_0$ the bubbles are classically expanding. 
During this process, the gain in the volume energy gets converted into kinetic energy of the bubble wall. 
Consider a set of expanding bubbles, that, at some initial time $t_0$, have zero kinetic energy $d_t\phi =0$. 
 
The energy of such a bubble $E_b$ is of course lower than the energy of the extremal static one, with $E(R_0)$.
The initial profile for such a bubble can in principle be found by equating \eqref{energyB} to $E_b$ and solving for $\phi$.
In practice, this might not be easy (or possible analytically), but there always exists a solution. 
Furthermore, we must remember that this solution only describes the bubble configuration $\phi(t_{0}, r)$ at some initial time $t_0$ with the initial condition  $d_t\phi =0$.
However, since $d_t^2 \phi \neq 0$, the bubbles evolve in time.  
  
A special case is given by the bubble $E_b = 0$. In the thin-wall limit, the radius of such a bubble can be easily found  by equating  the expression \eqref{energy(R)} to zero, which gives, 
\begin{align}  \label{Rzero} 
    R_0 = 3\frac{\sigma}{\epsilon} \sim \frac{m}{M^2}. ~~ 
\end{align}  
The special feature of the zero-energy bubbles is that they can be created without any energy transfer from the thermal bath.
This will be discussed in detail later on. 
   
\subsubsection{Oscillatory bubbles} 

If at some initial time the bubble has zero kinetic energy and $R < R_0$, it will oscillate.
"Oscillations" mean that the bubble radius is expected to shrink and expand several times before the final decay. 
During this process some particle production will take place. 
The duration of the oscillation period and stability with respect to small non-spherical perturbations is an open question. 
To our knowledge, the problem has not been solved fully.

Qualitative arguments suggest that for an oscillating thick wall bubble the lifetime should be prolonged by a factor $\sim \frac{1}{\alpha m}$~\cite{Saturon2020}. 
In this case the bubble can be viewed as a type of oscillon~\cite{Bogolyubsky:1976nx, Bogolyubsky:1976sc, Gleiser:1993pt, Kolb:1993hw, Copeland:1995fq} (for a recent analysis, see \cite{vanDissel:2025xqn} and references therein).

Notice that oscillatory bubbles exist for arbitrarily small values of $M^2(T)$ and can be of both types: thick-wall and thin-wall.     
Obviously, such bubbles have non-zero energy. 
In the thin-wall limit the energy is related to its initial radius via \eqref{energy(R)}. 
In the thick-wall regime, the average radius and the energy of oscillatory bubbles are given, order of magnitude wise, by
\begin{align}  \label{ERoscTW} 
    R \, \sim \,  \frac{1}{m}, ~~ 
    E_{\rm b} \, \sim \, \frac{m}{\alpha} \,. 
\end{align}  

Notice that a collapsing bubble can be stabilized by the $O(N)$ Goldstone charge \cite{Dvali:2021tez,Dvali:2024hsb}, which can be interpreted as a manifestation of the memory burden effect \cite{Dvali:2018xpy, Dvali:2020wft, Dvali:2025sog}.

\subsection{Constraints from the Species Scale}

Eventually, we are going to treat the colliding bubbles as the sources of gravitational waves. 
However, in our discussion the back-reaction of gravity on the bubble dynamics shall be ignored to leading order.  
This is a justified assumption under the conditions that the scale of the bubble wall tension is well below the Planck mass~\cite{Vilenkin:1984ib}.      
This assumption is rather common in the conventional analysis of first order phase transitions.   
  
However, in the present case the following gravitational constraint must be taken into account.  
Our main novelty lies in appreciating the effect of a microstate degeneracy on the phase transition. 
In our prototype model, this degeneracy originates from the internal $O(N)$-symmetry.  
Although gravity is blind with respect to this symmetry, it nevertheless imposes a very important constraint on $N$, since this quantity controls the number of particle species in the theory.  
Correspondingly,  the non-perturbative gravitational cutoff is lowered to a so-called species scale ~ \cite{Dvali:2007hz, Dvali:2007wp}, 
\begin{equation} \label{species} 
     \Lambda_{\rm gr} = \frac{M_P}{\sqrt{N}} \,.
\end{equation} 
Thus, for the QFT description to remain valid,  the gravitational cutoff $\Lambda_{\rm gr}$ must be above the QFT cutoff $\Lambda$. 
It is easy to see that this constraint is fully compatible with all the above-discussed double-scaling regimes, \eqref{DSlimit}, \eqref{TNlimit} and \eqref{CommonL}. 
In what follows, we shall assume that it is satisfied.  
   
\subsection{Microstate Degeneracy} 
 
We shall now discuss the microstate degeneracy of a bubble.
In order to do so, the microscopic count of \cite{Saturon2020} shall be used as reference. 
The origin of the microstate degeneracy of the bubble can be understood
from several perspectives. 

The interior of the bubble corresponds to the vacuum with a spontaneously-broken $O(N)$-symmetry, which supports $N-1$ gapless Nambu-Goldstone modes.
In other words, the bubble realizes the mechanism of assisted gaplessness, in which the Goldstones play the 
role of the memory modes \cite{Dvali:2018xpy, Dvali:2020wft}. 
The diversity of these Goldstone excitations form an exponentially large number of degenerate microstates. 

From the point of view of the external vacuum, a classical bubble represents a state transforming under an exponentially large representation of the $O(N)$-group.
The number of microstates is given by the dimensionality of this representation.

Due to the $O(N)$-invariance of the Lagrangian, a classical bubble can be oriented along any unit vector $e_j$, in flavor space. 
Without loss of generality, this vector can be represented as a fixed vector $(1,0,0, ..., 0)$ rotated by an orthogonal matrix $O_{jk}(\Omega)$, parameterized by a set of angles $\Omega$, 
\begin{align}
  e_j(\Omega) = O_{j1}(\Omega). 
\end{align}
In this parameterization, each classical solution can be labeled by $\Omega$, as $\phi_j(x,\Omega)$.

In general, in the quantum picture, a classical soliton solution must be resolved as the expectation value of a quantum field over a corresponding coherent state (for an incomplete list of references, see~\cite{Dvali:2011aa, Dvali:2013eja, SolCoh1, Dvali:2017eba, 
S1, S2, Saturon2020, S4, Berezhiani:2024pub, vanDissel:2025xqn}).

In the present case the quantum field is $\hat{\phi}_j$ with the mode expansion,  
\begin{align}
    \hat{\phi}_j \, = \, \sum_{\vec{p}} \frac{1}{\sqrt{2V\omega_{\vec{p}}}}(e^{i\vec{p}\vec{x}}   \hat{a}_j(\vec{p}) \, + \, e^{-i\vec{p}\vec{x}}   \hat{a}^{\dagger}_j(\vec{p}) )\,.  
\end{align} 
Here $V$ is the volume and $\hat{a}_j(\vec{p}), \ \hat{a}^{\dagger}_j(\vec{p})$ are  the standard annihilation and creation operators of modes labeled by momentum $\vec{p}$  and frequency $\omega_{\vec{p}}$.  
The classical vacuum bubble represents the expectation value of the quantum field $\hat{\phi_j}$ over the coherent state~\cite{Saturon2020} 
\begin{align} \label{cohBB}
   \ket{\Omega}  =   {\rm exp} \left( \int_{\vec{p}} \sum_j \sqrt{N_{j,\vec{p}, \omega}}
 (\hat{a}^{\dagger}_j(\vec{p}) - \hat{a}_j(\vec{p})) \right)   \ket{0} \,, 
\end{align}
where the coherent state parameters $N_{j,\vec{p}, \omega}$ represent the mean occupation numbers of the corresponding momentum and flavor modes. 
 
Using the standard QFT canonical commutation relations, the above expression can be written in the following form (see e.g., \cite{Coh1}), 
\begin{align} \label{cohJ}
   \ket{\Omega}  =   {\rm exp} \left(-i \int \mathrm{d}^3x \sum_j(\phi_j(\Omega) \hat{\pi}_j   - 
   \pi_j(\Omega) \hat{\phi}_j) \right)  \ket{0} \,, 
\end{align}
where $\hat{\pi}_j(x)$ is the conjugated momentum and $\pi_j(x, \Omega)$ is its classical counterpart evaluated on the bubble solution $\phi_j(x,\Omega)$.  
Of course, the operators satisfy the standard commutation relations, e.g., $[\hat{\phi_j}(x), \hat{\pi}_j(x')] = i \delta^3(x-x')$. 
It is clear that, 
\begin{align} \label{BEV}
   \bra{\Omega}  \hat{\phi_j}(x) \ket{\Omega} \, = \, \phi_j(x, \Omega)\,.
\end{align} 
This expression also establishes the dictionary between the occupation numbers of individual modes $N_{\vec{p},j, \Omega}  = \bra{\Omega} \hat{a}^{\dagger}_j(\vec{p}) \hat{a}_j(\vec{p}) \ket{\Omega}$, and Fourier harmonics of the classical field
\begin{align} \label{Ftrans}
\phi_j \, = \, \sum_{\vec{p}} \frac{1}{\sqrt{2V\omega_{\vec{p}}}}(e^{i\vec{p}\vec{x}}   \alpha_j(\vec{p},\Omega) \, + \, e^{-i\vec{p}\vec{x}}   
\alpha^{*}_j(\vec{p}, \Omega) )\,,  
\end{align} 
 as 
\begin{align} \label{occupationflavor} 
\alpha_j(\vec{p},\Omega)\alpha^{*}_j(\vec{p}, \Omega)  =  N_{\vec{p},j, \Omega}  \,.  
\end{align} 
Summing over all momentum and flavor modes gives the total occupation number, 
\begin{equation} \label{occupationtotal} 
    \sum_j \sum_{\vec{p}} N_{\vec{p},j, \Omega}   = \, N_{\phi} \,. 
\end{equation} 
Notice that, due to $O(N)$-symmetry, the total occupation number 
$N_{\phi}$ is independent of $\Omega$.  
In particular, this follows from the orthogonality of the matrix $O_{jk}(\Omega)$.  
That is, all $O(N)$-rotated bubbles have the same total occupation numbers. 

Notice that the constituents of the bubble are off-shell relative to their asymptotically free counterparts \cite{SolCoh1}.
For the thick wall bubbles, $R \sim \delta$, which are our main focus, this off-shellness is only order one. This means that 
the free-particle relation between the wavelength (momentum) and 
frequency holds order of magnitude 
wise for the bubble constituents.   

The off-shellness plays no role in the degeneracy count, which comes purely from symmetry considerations. 
Nevertheless, the mild off-shellness helps to visualize the degeneracy, since the constituents can mentally be approximated by free quanta.  

The number of degenerate bubbles can be calculated by counting the number of all possible distributions of the total occupation number $N_{\phi}$ among $N$ distinct components $\phi_j$ \cite{Saturon2020}. 
For the purpose of counting, the bubble vacuum (moduli space) can be described by an effective Hamiltonian of the following form~\cite{S1}, 
\begin{equation} \label{Hmoduli}
    \hat{H} = X\left (\sum_{i=1}^N\hat{n}_i - N_{\phi} \right )\,, 
\end{equation}
where $X$ is a Lagrange multiplier and $\hat{n}_j$ are the number operators of $N$ distinct modes (moduli) parameterizing the bubble vacuum.

At large $N_{\phi}$, each distribution corresponds to a classical bubble with distinct orientation $\Omega$. 
The number of orthogonal states is given by the following binomial coefficient, 
\begin{align}\label{nstNA}
    \nonumber n_{\rm st} &= \frac{(N_{\phi}+N)!}{N_{\phi}!N!} \\
    &\simeq \sqrt{\frac{N+ N_\phi}{NN_\phi}}\left(\frac{N}{N_\phi}\right)^{N-N_0}\times \\
    \nonumber& \times\left(1+\frac{N}{N_\phi}\right)^N\left(1+\frac{N_\phi}{N}\right)^{N_\phi},
\end{align}
where Stirling's approximation has been used, assuming both $N$ and $N_\phi$ are large.
The maximal value of $N$ is given by Eq. \eqref{Ubound}. 
Furthermore, for a bubble in the thick wall regime, the number $N_\phi$ of $\phi$ quanta can be estimated as \cite{Saturon2020}
\begin{align}
    N_\phi \sim \frac{1}{\alpha}.
\end{align}
Hence, $N\sim N_\phi\sim 1/\alpha$. In this case, the entropy can simply be expressed as 
\begin{align} \label{AlphaB}
    S = \ln(n_{\rm st}) \simeq \frac{1}{\alpha} \,.
\end{align}
Thus, the entropy matches the expression \eqref{Alpha} to leading order, with negligible corrections.

Notice that for a thick bubble, $R \sim m^{-1}$, the maximal entropy  also satisfies, 
\begin{align} \label{AreaB}
    S  \sim (R f)^2 \,,
\end{align}
which saturates the area-law bound \eqref{Area}, since the scale of Poincar\'e breaking for such a bubble is $f$. 
The number of resulting microstates is correspondingly, 
\begin{align} \label{NstApp}
    n_{\rm st}  = {\rm e}^{S_{\rm sat}} \sim {\rm e}^{1/\alpha} \sim  {\rm e}^{N}
\end{align} 

Following \cite{Saturon2020}, the above degeneracy count can easily be generalized to thin-wall bubbles, which have $R \gg m^{-1}$.  
As shown there, the entropy of these bubbles is larger only by a factor $\sim \ln(Rm)$, 
\begin{align} \label{SthinR}
    S_{R \gg m^{-1}} \sim   \frac{1}{\alpha} \, \ln(Rm) \,. 
\end{align}                  
This growth can be interpreted as the ``running" of the effective coupling with the scale $R$.

\subsection{Bubble Nucleation} 
 
The goal now is to estimate the rate of the decay of the symmetric vacuum. 
This decay proceeds via bubble nucleation. 
The task is to identify the most probable bubble nucleation process.  
This shall be discussed in various regimes. 
 
\subsubsection{Bubble Nucleation in the Decoupling Limit} 
 
To start with, we take the exact double scaling limit \eqref{DSlimit}.  
In this limit the  $\chi$ sector fully decouples from the $\phi$ field. 
Thereby, no energy transfer from the thermal bath to $\phi$ is possible.
The situation in the  $\phi$-sector is fully equivalent to the zero temperature case with  a ``hard" correction to the mass given by $M^2(T)$.  
  
In this case, tunneling proceeds as in a zero-temperature false vacuum decay~\cite{Kobzarev, Coleman:1977py}, via nucleation of a  critical bubble of zero energy. 
It is therefore possible to follow the standard zero-temperature procedure, as laid out in~\cite{Coleman:1977py}.  
The new input is the microstate degeneracy. 
   
For a bubble with a particular $e_j$, the Euclidean trajectory leading to the tunneling process is dominated by the $O(4)$-invariant configuration $\phi(r)$, which satisfies    
\begin{align} \label{Bounce}
    \mathrm{d}_r^2 \phi_j(r) + \frac{3}{r}\mathrm{d}_r \phi_j(r) - \frac{\partial V}{\partial \phi_j} = 0 \,,    
\end{align} 
where $r$ must be understood as the radial coordinate in four-dimensional Euclidean space. 
The boundary conditions are such that $\phi(r)$ interpolates from $\phi(0) \neq 0$  to $\phi(\infty) = 0$.  As is well-known \cite{Coleman:1977py}, such a solution to \eqref{Bounce} always exists.  
The corresponding action is,    
\begin{align}  \label{Action4} 
    A_4  \, = \, 4\pi^2 \int_0^{\infty} r^3\mathrm{d}r \left ( \frac{1}{2} (\mathrm{d}_r \phi_j)^2 +  V(\phi_j) \right )\, . 
\end{align} 
 
\begin{figure}
     \centering
     \includegraphics[width=\linewidth]{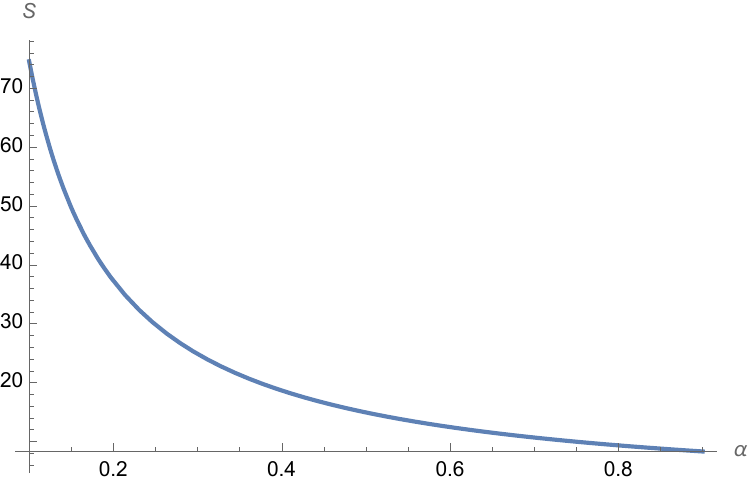}
     \caption{A plot of the Euclidean action $A_4$ for different values of $\alpha$ in the decoupling limit, with $M\sim m$. The action scales exactly as $1/\alpha$. The action has been calculated using the Mathematica package \textit{FindBounce}~\cite{Guada:2018jek, Guada:2020xnz}, taking $f\equiv 1$ and $M/m =0.9$.}
     \label{fig:decoupling limit}
\end{figure}

The nucleation rate of a bubble with a particular orientation of $\Omega$ is exponentially suppressed as,
\begin{align}  \label{Gj} 
  \Gamma_{\Omega} \, \sim {\rm e}^{-A_4}\,.  
\end{align} 
In the thin-wall limit, the derivation of $A_4$ simplifies to
\begin{align}  \label{A4R} 
    A_4(R) \, = \, \pi^2 \left ( - \frac{1}{2} R^4 \epsilon  + \sigma R^3 \right )\,,  
\end{align} 
where $\sigma$ is the ``tension" of the bounce wall.  
The critical radius, $R_0$, can be found by extremizing the above expression. This gives, 
\begin{align}  \label{A4R0} 
    R_0 = 3\frac{\sigma}{\epsilon} \sim \frac{m}{M^2}, \, A_4 = \frac{27\pi^2}{2} \frac{\sigma^4}{\epsilon^3} \sim \frac{1}{\alpha} \frac{m^6}{M^6}. 
\end{align}  
It can be seen that the action $A_4$ is lower for thick-wall critical bubbles. These exist in the regime $M \sim m$. Correspondingly,
the action satisfies, 
\begin{equation}  \label{Opt4} 
    A_4  \, \sim \, \frac{1}{\alpha} \,. 
\end{equation} 
Numerical calculations, done using the Mathematica package {\it{Find Bounce}}~\cite{Guada:2018jek, Guada:2020xnz}, confirm this behavior, see Fig. \ref{fig:decoupling limit}. 
These results were obtained by calculating the action for different values of $\alpha <1$ for $M=cm$, with $c\lesssim 1$ ($c=0.9$ in Fig. \ref{fig:decoupling limit}), taking $f\equiv 1$.

The total nucleation rate is obtained by summing $\Gamma_{\Omega}$ over all degenerate microstates. 
Due to the $O(N)$ symmetry, this amounts to integration over the set of angles $\Omega$, or equivalently to multiplying the rate by the number of microstates \eqref{NstApp}: 
\begin{equation}  \label{GT4} 
    \Gamma = \int \mathrm{d}{\Omega} \ \Gamma_{\Omega} \sim {\rm e}^{-A_4 + S}
\end{equation}  

As is clear from the expressions \eqref{NstApp}  and \eqref{Opt4}, if the microstate degeneracy of the bubble saturates the bound \eqref{Alpha} (equivalently \eqref{Area}), the entropy factor of the bubble can overcome the exponential suppression.
That is, the vacuum bubble and the corresponding Euclidean bounce saturate the unitarity bounds 
\eqref{Alpha} and \eqref{Area} simultaneously. 
Due to this, the bubble nucleation can proceed unsuppressed, suggesting that beyond the saturation point the theory enters a new regime. 

 As originally discussed in \cite{S2}, the phenomenon of the regime-change, which takes place upon saturation of the entropy bounds \eqref{Alpha} and \eqref{Area} by an instanton, is generic.
In this work, it was shown that the saturation of the bounds \eqref{Alpha} and \eqref{Area} by a Yang-Mills gauge instantons is correlated with the theory entering the confinement regime. 
 
Notice that, since the bubble has zero energy (relative to the energy of the initial thermal bath), it is not possible to constrain its entropy using the Bekenstein bound \eqref{Bek}. 
This would miss the correct degeneracy given by \eqref{NstApp}. 

Instead, the entropy must be constrained by the QFT bounds \eqref{Alpha} and \eqref{Area}, which work even in cases for which the Bekenstein bound cannot be defined. 
The same consideration applies to the instantons \cite{S2} and in the present case to the Euclidean bounce, for which the energy cannot be defined.

\subsubsection{The issue of boost integration}

In the regime in which the zero energy bubbles are meterialized, the following issue must be clarified. 
It was already pointed out that in the strict decoupling limit (\ref{DSlimit}) there is no energy transfer between the thermal bath and the bubble.  
Due to this, the nucleation process resembles false vacuum decay at zero temperature.  

It was pointed out in~\cite{Dvali:2011wk} that, if the false vacuum is a zero-temperature Minkowski state, the tunneling creates the issue of integration of the created bubble over the Lorentz boosts. 
Namely, since the bubble has zero energy, the boost integration results in an infinite nucleation rate. 
This implies that the would-be ``false vacuum", in reality, does not exist as a well-defined state in the Hilbert space.  

The paper~\cite{Dvali:2011wk} concludes that this divergence signals that in a consistent theory a tunneling from the Minkowski vacuum cannot happen.  
This statement fully agrees with arguments made earlier by  Zeldovich in~\cite{Zeldovich:1974py}. 

Notice that, the boost-invariance of the ``eternal" Euclidean bounce~\cite{Coleman:1977py} does not provide a loophole in this statement~\cite{Dvali:2011wk}. 
First, viewed as an $S$-matrix process, the nucleation of a classical bubble $\ket{bubble}$
with a particular boost factor, breaks Poincar\'e  symmetry. 
Therefore, the transition must take place into a boost-invariant superposition of bubbles integrated over the entire Poincar\'e group. 
\begin{equation} \label{POinv}
  \ket{\rm inv}_{\rm Poincare} =  C \int_{\rm boost} \ket{\rm bubble} \,, 
\end{equation}
where $C$ is a normalization factor. 
If the transition matrix element, 
$\bra{\rm bubble} {\rm e}^{-i\hat{H}t} \ket{\rm false}$,
from initial false vacuum to a particular classical bubble is finite, the transition to  the boost-integrated bubble  
state (\ref{POinv}) gives an infinite factor. 
  
Second, one can consider the infinitesimal Poincar\'e-violating deformations that make the boost-integration necessary, and then remove the deformation recovering the infinite rate. 
Moreover, one may argue that meta-stability of the state makes the question of the Poincar\'e invariance of the bounce irrelevant.   
 
The paper~\cite{Dvali:2011wk} further states that in gravity, the energy splitting between Minkowski and any lower energy vacuum must be below the Coleman-De Luccia bound~\cite{Coleman:1980aw}, which guarantees the absence of tunneling. 
That is, in the flat space limit, the vacua must be strictly degenerate. 

This view is independently strengthened by the argument that the consistent vacuum of quantum gravity is Minkowski, excluding any type of a metastable de Sitter \cite{Dvali:2013eja, Dvali:2014gua, Dvali:2017eba}. 
This requirement, in particular, is enforced by the $S$-matrix formulation of quantum gravity~\cite{Dvali:2020etd}. 
This picture demands that the asymptotic vacua of quantum gravity are Minkowski, whereas the non-trivial gravitational backgrounds, such as 
de Sitter space, must be described as graviton 
coherent states constructed on top of the Minkowski vacuum \cite{Dvali:2011aa, Dvali:2013eja, Dvali:2014gua, Berezhiani:2016grw, Dvali:2017eba}.
 The BRST-invariant construction of such 
 states can be found in~\cite{Berezhiani:2021zst, Berezhiani:2024boz}. 

 We must also remark that, interestingly, the Minkowski criterion of~\cite{Dvali:2011wk} supports the so-called principle of ``multiple point criticality''~\cite{Bennett:1993pj, Bennett:1996vy}. 

Despite the importance of the above issue, in our discussion of tunneling in the decoupling limit (\ref{DSlimit}), the problem of integration over boosts raised in~\cite{Dvali:2011wk} shall be put aside. 
This is justified due to the following reason. 
The decoupling limit is an idealized approximation, which cannot be exact in realistic situations. 
Since the thermal bath breaks the Poincar\'e symmetry, regardless how weakly it couples to the bubble field, a bubble boosted relative to the bath will acquire a non-zero energy. Due to this, the boost integration will be effectively cut off.
Of course, the residual integral shall enhance the rate by a finite factor, which in quantitative analysis must be taken into account. 
However, we wish to clearly separate this enhancement from the effects of the internal $O(N)$-degeneracy we are after.  
Therefore, in what follows, the effects from the boost integration relative to the thermal bath shall be ignored, bearing in mind that this will give an additional, $O(N)$-independent factor. 

\subsubsection{Common Saturation Couplings} 
 
Now, the regime described by equation \eqref{CommonL}, where 
\begin{align}\label{ABsaturation} 
    \beta \sim \alpha \sim 1/N\,, 
\end{align}  
will be further investigated.
Both couplings $\alpha$ and $\beta$ saturate their unitarity bounds, given by eq. \eqref{Ubound} and eq. \eqref{BetaB}.  
However, since $N$ is large, both couplings are nevertheless very weak.   
In this regime, even though the $\phi$'s are not in thermal equilibrium, the thermal bath strongly changes the dynamics of bubble-nucleation, provided the bubble radius satisfies  $R \gg T^{-1}$. 
  
In this regime, the most relevant bubble configurations describing the decay of the false vacuum aren't zero energy bubbles, but extremal static bubbles, obtained by solving the equation \eqref{eqST}. 

This can be understood from the following reasoning: 
QFT at finite temperatures is mapped to a theory with a compactified temporal direction, with radius $1/T$. 
For $R \gg T^{-1}$, the least costly Euclidean  trajectory corresponds to a $3$-dimensional Euclidean bounce with action $A_3$ \cite{Linde}. 
This bounce is obtained by solving the equation 
\begin{align} \label{B3}
    \mathrm{d}_r^2 \phi_j(r) + \frac{2}{r}\mathrm{d}_r \phi_j(r) - \frac{\partial V}{\partial \phi_j} = 0 \,.   
\end{align} 
where $r$ is the Euclidean radial coordinate. 
  
This equation is identical to \eqref{eqST}.  Hence the solution formally coincides with a static extremal bubble in $3+1$-dimensional Lorentzian space-time. The Euclidean action of the bounce  is given by the integral, 
\begin{align}  \label{A3} 
    A_3  \, = \, 4\pi \int_0^{\infty} r^2\mathrm{d}r \left ( \frac{1}{2} (\mathrm{d}_r \phi_j)^2 +  V(\phi_j) \right )\, , 
\end{align} 
which coincides with the energy  \eqref{energyB} of a static Lorentzian bubble. 
 
In particular, analog to \eqref{energy(R)}, utilizing the thin-wall approximation gives, 
\begin{align}  \label{A3R1} 
    R_0 = 2\frac{\sigma}{\epsilon} \sim \frac{m}{M^2}, ~~ A_3 = \frac{16\pi}{3} \frac{\sigma^3}{\epsilon^2} 
  \sim \frac{m}{\alpha} \frac{m^4}{M^4} \,. 
\end{align}  
In other words, $A_3$ represents the Euclidean action of the bounce, creating a static bubble of the $\phi$-field from the thermal bath of $\chi$-particles.
   
\begin{figure}
     \centering
     \includegraphics[width=\linewidth]{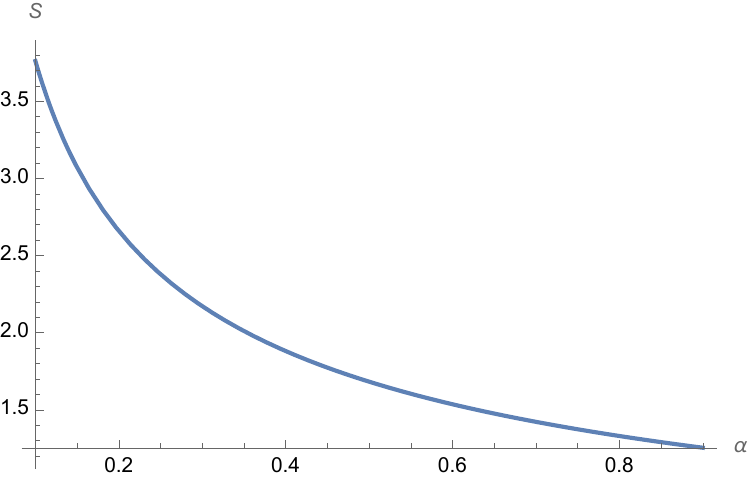}
     \caption{A plot of the thermal action $A_3/T$, for different values of $\alpha$ in the exact common saturation limit. The action possesses an exact $1/\sqrt \alpha$ scaling. The action has been calculated using the Mathematica package \textit{FindBounce} \cite{Guada:2018jek, Guada:2020xnz}, taking $f\equiv 1/\sqrt{\alpha}$, $\beta =0.9\alpha$ and $T=1/\sqrt\alpha$.}
     \label{fig:common limit}
 \end{figure}

Comparing expressions \eqref{A4R0} and \eqref{A3R1}, it is easy to see that if $R \gg T^{-1}$, then
\begin{align}  \label{A3R2} 
    A_4 \sim A_3 R \gg \frac{A_3}{T} \,. 
\end{align}  
Correspondingly the total nucleation rate \eqref{GT4} changes to, 
\begin{align}  \label{GT3} 
    \Gamma = \sum_{\Omega} \Gamma_{\Omega} \sim {\rm e}^{-\frac{A_3}{T} + S} \,.
\end{align}   
Notice that in the given parameter space, the saturated static bubbles exist at temperatures $T \sim m/ \sqrt{\alpha}$.  
This is  because  $m\sim M \sim \sqrt{\alpha} T$. 
Then, taking into account \eqref{A3R1}, this leads to,  
\begin{align}  \label{A3Tsat} 
  \, \frac{A_3}{T} \, \sim    \frac{1}{\sqrt{\alpha}}  \,,  
\end{align}  
for these type of bubbles. 
This is much smaller than the entropy factor $S_b \sim \frac{1}{\alpha}$.
Correspondingly, the entropy factor wins and the transition proceeds unsuppressed, signaling that the theory enters a new regime. 
Just like in the decouling regime, this can be confirmed in the limit, using numerical calculations, see Fig. \ref{fig:common limit}

However, we notice the following two things. 
First, the entropy of a bubble of radiation, with radius $1/m$ is 
\begin{align}
    S_{rad} \sim (T/m)^3 \sim 
  1/\alpha^{3/2} \sim \frac{1}{\sqrt{\alpha}}S_{\rm b}.
\end{align}
Therefore the entropy of the "radiation bubble" is larger than the entropy of the vacuum bubble, by a factor $1/\sqrt{\alpha}$.
However, its energy is 
\begin{align}
    E_{\rm rad} \sim T^4/m^3 \sim 
  m/\alpha^{2} \sim E_{\rm b}/\alpha \gg E_{\rm b}.
\end{align}
Thus, the thermal bath ``invests" only the fraction $\sim \alpha$ of its energy into the creation of the bubble.  
However, this portion of energy accounts only for $\sim \sqrt{\alpha}$ of the bubble entropy. 
Thus, since only a fraction of the radiation energy had to be invested, the entropy of this invested portion has been increased by a factor of $1/\sqrt{\alpha}$. This is because the invested energy is $\Delta E_{\rm rad} \sim \alpha E_{\rm rad}$ and therefore $\Delta S_{\rm rad} \sim \alpha S_{\rm rad}$. 
This can be used to obtain, 
\begin{align} \label{DerS}
   - \frac{\Delta S_b}{\Delta S_{rad}} \sim N^{\frac{1}{2}} \,. 
\end{align} 
This quantity measures the relevant gain in entropy during the bubble nucleation.

\subsubsection{Nucleation of Oscillon Bubbles} 
 
Just as before, the limit given by equation \eqref{CommonL}, where $\alpha \sim \beta \sim 1/N$, is assumed to hold.  
 We assume further that the $A_3/T$ factor is much larger than the bubble entropy.
Taking into account the expressions \eqref{A3R2} and \eqref{SthinR}, this implies, 
\begin{align}\label{Tosc} 
    T \ll \frac{m}{\alpha^{\frac{2}{5}}} \,.
\end{align} 
Thus, the  suppression factor $exp(-A_3/T)$ cannot be compensated by the bubble entropy. Therefore, the transition through the nucleation of thermal bubbles is exponentially suppressed. 
 
Nevertheless, in this situation unsuppressed transitions can still occur through the nucleation of a different type of saturated bubbles: an oscillatory one.  
In the regime of equation \eqref{ABsaturation}, an oscillatory saturon bubble exists for arbitrarily small values of $M < m$. 
Following \cite{SaturonDM}, it will be shown that it can be created at an unsuppressed rate at a certain optimal temperature $T_*$. 
     
As already discussed, an oscillatory bubble has a radius $R$, smaller than the radius $R_0$ of the static bubble given by the equation \eqref{energy(R)}. 
The specific interest here lies in saturated ones, satisfying $R \sim 1/m$.     
  
In order to derive $T_*$, consider the energy and entropy densities of $\chi$ radiation, which are given by the standard expressions,  
\begin{align} 
    \rho_{\rm rad} = \frac{\pi^2}{30}  T^4\, ~~ {\rm and}~~  s_{\rm rad} =  \frac{2\pi}{45}  T^3 \,.
\end{align} 
This can be translated into a relation between the energy and the entropy of a bubble of radiation, with radius $R$
\begin{align} \label{SandErad}
    S_{\rm rad} =  \frac{4}{3 \pi}\, \frac{E_{\rm rad}}{T}.
\end{align} 
Now, we demand that the saturon and the radiation bubbles have the same energy, 
\begin{align} \label{EradEsat2} 
    E_{\rm rad } \, = \,  E_{\rm bub} \, \sim \frac{m}{\alpha}\,.  
\end{align} 
This gives the condition 
\begin{align} \label{relationE2} 
    T_* \sim  m \frac{1}{\alpha^{\frac{1}{4}}}.
\end{align} 
At this temperature, the entropy of a saturated oscillon bubble with radius $R\sim 1/m$ exceeds the entropy of a bubble of radiation with the same radius and energy. 
Therefore, it is expected that the thermal formation of these saturons takes place unsuppressed. 
Indeed, applying equation \eqref{SandErad} at $T=T_*$ one obtains, 
\begin{align} \label{Sradlim} 
     S_{\rm rad}(T_*) \, \sim \frac{1}{\alpha^{\frac{3}{4}}} \ll S_{\rm bub} \sim \frac{1}{\alpha} \,.
\end{align} 
 Thus, from the entropy point of view,
it is profitable to convert the thermal bath of $\chi$-particles into saturated oscillatory bubbles of $\phi$-s.
 
As already explained in \cite{Saturon2020}, from the QFT perspective, it is useful to think about the transition process as a multiparticle scattering process, in which the initial state of $\chi$ quanta, with occupation number density $n_{\chi} \sim S_{rad}$ $T_*^3$ and characteristic momentum $\sim T_*$, transitions into the coherent state of $\phi$-s with $N_{\phi} \sim 1/\alpha \sim  \frac{S_{rad}}{\alpha^{\frac{1}{4}}}$. 
Due to the large difference in occupation numbers, the transition probability in each particular microstate of $\phi$-s is suppressed as $exp(-1/\alpha)$.
This factor is compensated by the high multiplicity of degenerate final states. 
This manifests the QFT interpretation of the saturon phase transition. 

However, the above outline is highly qualitative 
and for a cleaner conclusion a more explicit calculation is required. 
We must also keep in mind, that the unsuppressed saturon transition 
is correlated with the regime-change of the theory, which must 
be understood on case by case basis. 

\section{Macroscopic Entanglement} 

We shall now discuss a qualitatively new feature brought on by the microstate degeneracy of the vacuum bubble. 
Namely, the degeneracy forces the bubble, no matter how macroscopic, to be materialized in a quantum state with maximal entanglement. 
In the present case, this entanglement can be understood entirely in terms of the $O(N)$-symmetry. 

Obviously, since the state $\ket{\Omega}$ and the corresponding field configuration $\phi_j(\Omega)$ have fixed orientations in $O(N)$ space, the bubble state $\ket{\Omega}$ is not $O(N)$-invariant.
This simply reflects that the classical bubble breaks the $O(N)$ symmetry spontaneously.

However, the bubble has to materialize in a state that is $O(N)$-invariant. This can be understood from the symmetry properties of the time evolution operator: The Hamiltonian of the system commutes with the generators of $O(N)$. 
Correspondingly, the time evolution preserves the $O(N)$ symmetry. 

Now, the initial state corresponding  to the $\phi = 0$ vacuum is $O(N)$-invariant,
whereas the state $\ket{\Omega}$, given by  \eqref{cohJ}, is not. 
This state describes a classical bubble of $O(N -1)$ vacuum, with a particular orientation $\Omega$ of the $\phi$-field,
and transforms under  $O(N)$ non-trivially.      
Such a state cannot be produced as a result of the time evolution of the symmetric one.

Instead, the bubble materializes in an $O(N)$-invariant state that 
represents a superposition of states like eqn. \eqref{cohJ} integrated over the entire group manifold, 
\begin{align}  \label{invB}
    &\ket{\rm {inv}} = C \int \d\Omega \ket{\Omega}  = \\  \nonumber  
    & = C \int \d\Omega  \exp\left({-i \int \d^3x \sum_j(\phi_j(\Omega)\hat{\pi}_j  - \pi_j(\Omega)\hat{\phi}_j)}\right)  \ket{0},
\end{align} 
where $C$ is a normalization factor. 

This superposition includes orthogonal states with highly entangled modes.
The number of distinct states entering the superposition grows exponentially with $N$. 

In order to understand this $N$-scaling, it
suffices to consider a simplified picture in which the bubble is entirely composed of modes with a particular momentum. 
This approximation is especially good 
for a thick-wall bubble, for which the radius satisfies $R\sim 1/m$.
Correspondingly, such a bubble is dominated by modes of wavelengths $R\sim m ^{-1}$ (i.e., of momenta $|\vec{p}| \sim 1/R \sim m$) with occupation number $N_{\phi} \sim 1/\alpha$ \cite{Saturon2020}.  
Therefore, for illustrative purposes, we can take $N_{\phi} \sim 1/\alpha$ as the mean occupation number for a coherent state with a single momentum mode.   
 
The $O(N)$-invariant coherent state composed of a singular mode $\hat{a}_j^{\dagger}$ of particular momentum $|\vec{p}| \sim 1/R$, with mean occupation number $N_{\vec{p}} = N_{\phi}$, is given by, 
\begin{align} \label{coh}
   \ket{{\rm inv}}  \, = \,C \int \d\Omega \,  {\rm e}^{-\frac{1}{2} N_{\phi}}
   {\rm e}^{ \sqrt{N}_{\phi} \left(\sum_j\alpha_j(\Omega)\hat{a}_j^{\dagger} \right)}  \ket{0}\,.
\end{align}
This state represents a superposition of invariant states of the type, 
\begin{align} \label{NCstates} 
(\sum_k\hat{a}_k^{\dagger}\hat{a}_k^{\dagger})^{n/2} \ket{0} , 
\end{align} 
with integer $-\infty \leqslant n\leqslant + \infty$. 
The maximal weight is carried by states with $n=N_{\phi}$.
Therefore, the invariant state $\ket{{\rm inv}}$ represents a superposition of orthogonal states, scaling exponentially with $N_\phi \sim N$. 
 
Another way to see this, is by observing that the superposition \eqref{coh} can always be written in terms of number eigenstates of the $j$-modes, $n_j = \langle \hat{a}_j^{\dagger} \hat{a}_j \rangle$, 
\begin{align} \label{sumMCR}
    \ket{{\rm inv}}  =  \sum_{n_1,...,n_N}  c_{n_1,...,n_N} \ket{n_1,...,n_N}    
\end{align} 
where $c_{n_1,...,n_N}$ are the respective binomial coefficients and the summation is subject to $\sum n_j = N_{\phi}$.
  
This counting also holds for invariant states \eqref{invB} that include all momentum modes.   
The difference is that in this case the summation goes over invariant states with all possible momenta contained in the Fourier harmonics of the classical field  $\phi_j(x)$.  

It is now easy to understand the role the microstate degeneracy plays for the quantumness of the bubble state. 
The $O(N)$-invariant bubble \eqref{invB} represents a superposition of would-be classical bubbles, described by \eqref{cohJ}. 
The number of classical participants grows exponentially with $N$ and is maximal for saturated bubbles, where $N = N_{\phi} = 1/\alpha$. 
 
In this case, the number of entangled states entering the invariant bubble is given by \eqref{NstApp}. 
Of course, as the bubble expands, the occupation number $N_{\phi}$ grows, whereas $N$ is fixed.  
Therefore, if an initially-saturated bubble grows substantially it becomes undersaturated.  
Regardless, the microstate degeneracy guarantees the quantumness of the bubble state.  

The phenomenon is present for arbitrary $N$, but the entanglement of the state is more resistant against external perturbations for larger values.  
The larger $N$ is, the less effective external perturbations are in classicalizing the state.   
For example, for $N\gg 1$,  measuring the state of one of the constituent modes, say  $j=1$, still leaves the system in an entangled superposition of an exponentially large number of states.
On average, it will take of order $N$ measurements (one per each $j$-flavor) to collapse the state into one of the basic classical states.  

In order to illustrate the effect of $N$-scaling, as an extreme example, let us consider the case of a $Z_2$ degeneracy, corresponding to $N=1$. 
In this case, the invariant bubble represents a $Z_2$-symmetric superposition of only two classical bubbles, 
\begin{equation} \label{Z2}
    \ket{{\rm inv}} =  \frac{1}{\sqrt{2}} (\ket{{\rm bubble}}_{+}  +   \ket{{\rm bubble}}_{-}). 
\end{equation} 
The states $\ket{\rm{bubble}}_{\pm}$ describe the classical bubbles that are related to each other by the $Z_2$ transformation, which 
acts on the field as $\phi \rightarrow -\phi$.
Of course, as in the case of large $N$, each classical bubble represents a coherent state of macroscopic occupation number $N_{\phi}$. 
However, in the $Z_2$ case, the superposition \eqref{Z2} consists of only two classical (i.e. coherent) states.  
While the modes in this superposition are highly entangled, the superposition is more "fragile" against the external perturbations as compared to the analogous 
$O(N)$-invariant state (\ref{invB}) with $N>1$. 

Fragility means in this case that for smaller $N$ it takes less external effort to project the entangled $O(N)$-invariant quantum state \eqref{invB} to one of its coherent constituents \eqref{coh}, which are essentially classical states.

Indeed, for arbitrary $N$, a classical bubble \eqref{coh} represents a coherent state of $N_{\phi}$ constituents. 
However, in case of a classical $Z_2$-bubble, either $\ket{{\rm bubble}}_{+}$ or $\ket{{\rm bubble}}_{-}$, 
the state of a single constituent fully identifies the bubble state.
Correspondingly, a single external measurement of the state of one of the constituent, suffices to project the $Z_2$-invariant state \eqref{Z2} to a corresponding classical bubble state.  

This is not the case for large $N$. 
Indeed, to reduce the $O(N)$-singlet state \eqref{invB} to a particular classical bubble state \eqref{cohJ}, one needs to measure the states of $\sim N$ constituents. 

\section{Entanglement Imprints in Gravitational waves}

The microstate degeneracy of the merging vacuum bubbles, has a macroscopic effect on gravitational radiation. 
This is due to a high level entanglement of the bubble states.  
 
In order to understand this effect, let us follow the process step by step. 
First, we consider the nucleation of an isolated bubble from an $O(N)$-invariant vacuum.  
As already discussed, due to the $O(N)$-invariance of the Hamiltonian, the bubble must materialize in an $O(N)$-singlet state \eqref{invB}. 
This represents a superposition of $O(N)$-non-invariant coherent states \eqref{cohJ} which, to a good approximation, describe classical bubbles with different orientations of the $\phi_j$-field.   
Thus, we can say that the $O(N)$-symmetry of the Hamiltonian forces the bubble to materialize in a highly quantum state of an $O(N)$-invariant bubble \eqref{invB} which represents a maximally entangled superposition of "classical" bubbles \eqref{cohJ}. 
    
At the moment of materialization, say $t=0$, the invariant bubble state can schematically be written as
\begin{align} \label{NB}
    \ket{{\rm inv}}_0 = \, C\, \sum_{A}  \ket{{\rm cl(0) }}_A\,, 
\end{align} 
where the states $\ket{{\rm cl(0)}}_A$ describe classical bubbles represented by coherent states of the type \eqref{cohJ}, where the index $A$ runs over their characteristics. 

The next stage is the process of bubble expansion. 
The bubble state is evolved in time by the action of the exact time-translation operator  ${\rm e}^{-i\hat{H}t}$, 
\begin{align} \label{QbubbleT}
    \ket{{\rm inv}}_t = \, C\, \sum_{A}  \ket{{\rm cl(t) }}_A\,, 
\end{align} 
where 
\begin{equation} \label{CLBT}
    \ket{\rm{cl}(t)}_A \equiv {\rm e}^{-i\hat{H}t}
    \ket{{\rm cl(0) }}_A \,,
\end{equation}
is the time-evolved state from an initially classical bubble
of label $A$.

The nature of the state \eqref{CLBT} requires some clarification. 
Ordinarily, the classical description of a time-evolving quantum system can break down after a certain critical time, called the \textit{quantum break time}. 
The concept was introduced in~\cite{Dvali:2013vxa}, where it was also shown that for unstable systems the quantum break time can scale logarithmically in the number of constituents. 
The phenomenon of quantum breaking has been further studied in various systems, using different methods~\cite{Dvali:2013eja, Dvali:2017eba, Berezhiani:2021zst, Berezhiani:2020pbv, Berezhiani:2021gph, Berezhiani:2023uwt, Berezhiani:2025tkp}. 
The quantum break time for expanding or merging classical bubbles has never been studied. 
In the present case, it would manifest in a departure of the state $\ket{\rm{cl}(t)}_A$ from the classical evolution. 
The latter essentially reduces to a classical expansion of the bubble. 
 
In particular, the departure from the classical evolution can be parameterized by the growth of the higher point functions of the $\hat \phi_j$-field.
If such quantum departures from the classical evolution of an isolated expanding bubble are significant, this will only strengthen our point about the importance of quantum effects. 
However, in order to cleanly isolate the quantum effects coming from the microstate degeneracy, which forces bubbles to be highly entangled, we shall consider the two cases separately. 

That is, we shall fist assume that 
the time evolution of an initially classical bubble would not violate classicality on the time scales of interest. 
Correspondingly, all of the quantumness comes from the superposition of individual classical trajectories, as opposed to their breakdown.
In other words, the true quantum evolution of the invariant bubble \eqref{QbubbleT} represents a superposition of classical trajectories of the expanding classical bubbles.  

The next stage of the time evolution is the merger. 
We shall assume that the bubble merges with another bubble, which was also materialized in an $O(N)$-invariant state 
\begin{align} \label{Bprime}
    \ket{{\rm inv}}_0' = \, C\, \sum_{B}  \ket{{\rm cl(0) }}_B\,. 
\end{align} 
Of course, prior to the merger, the time evolution of the partner bubble is similar, modulo the formation time. 

Now, when the bubbles are far apart, the state can be represented by the tensor product 
\begin{equation} 
    \ket{\rm in} =  \ket{{\rm inv}}_0 \otimes \ket{\rm{inv}}_0' \,,
\end{equation}
and is further evolved in  
time as, 
\begin{equation} \label{QtEV}
    \ket{t} = {\rm e}^{-i\hat{H}t}\ket{\rm in} = \, C\, \sum_{A,B} \ket{{\rm cl(t) }}_{A,B} \,,
\end{equation}
where 
\begin{equation} \label{CLtEV}
    \ket{\rm{cl}(t)}_{A,B} \equiv {\rm e}^{-i\hat{H}t}\ket{\rm{cl}(0)}_A \otimes \ket{{\rm cl(0) }}'_B \,.
\end{equation}
Of course, after $t> t_{\rm merger}$
the state $\ket{\rm{cl}(t)}_{A,B}$ is no longer a tensor product, since it describes the classical state obtained by the merger of two classical bubbles. 
However, assuming the quantum breaking of individual classical trajectories remains small, the state $\ket{\rm{cl}(t)}_{A,B}$ is still approximately classical. 
That is, the true time-evolving quantum state of the merging bubbles represents a superposition 
of $n_{st} \times n_{st}$  classical trajectories, each describing a merger of a pair of classical bubbles with characteristics $A,B$. 
 
We thus see that the merger of true quantum bubbles is macroscopically different from the mergers of would-be classical ones. 
This can have a significant effect on gravitational radiation. 

In order to see this, let us consider the effective equation describing classical gravitational radiation,    
\begin{equation} \label{gravity}
   \Box \, h_{ab} = 16\pi G \bra{t} \hat{T}_{ab} \ket{t} \,,   
\end{equation}
where $h_{ab}$ are the space components of a transverse-traceless linear gravitational field and $\hat{T}_{ab}$ is the energy momentum tensor of the field $\hat{\phi}$.
It is clear, that the expectation value taken over the state $\ket{t}$, given in \eqref{QtEV}, is different from the analogous expectation values taken over any of the states $\ket{\rm{cl}(t)}_{A,B}$ describing the would-be classical mergers and, instead, represents their sum,   
 \begin{equation} \label{Tvev}
    \bra{t} \hat{T}_{ab} \ket{t} = \sum_{A,B} T_{ab}^{(A,B)} \,,
 \end{equation}
where $T_{ab}^{(A,B)} = \, |C|^2\, \bra{\rm{cl}(t)}_{A,B}\hat{T}_{ab}\ket{\rm{cl}(t)}_{A,B}$.

Notice that, due to classicality of the states $\ket{\rm{cl}(t)}_{A,B}$, the off-diagonal matrix elements $\bra{\rm{cl}(t)}_{A,B}\hat{T}_{ab}\ket{\rm{cl}(t)}_{A',B'}$, with $A\neq A'$ and $B \neq B'$ can, to leading order, be set to zero. 

In other words, even when we take a conservative approach and assume the validity of the classical approximation in a classical merger, we find that the merger of $O(N)$-invariant bubbles represents a quantum superposition of an exponentially large number of very different classical trajectories. 

Correspondingly, the resulting \textit{classical} gravitational wave  represents the linear superposition of gravitational waves emitted by an exponentially large number of classical mergers, 
\begin{equation} \label{waveAB}
    h_{ab} = \, |C|^2\, \sum_{A,B} h_{ab}^{(A,B)} \,.
\end{equation} 
Due to $O(N)$-symmetry, many classical mergers would give identical gravitational waves. 
But in particular, mergers giving distinct gravitational waves can be parameterized by the relative angle $\theta$ between the expectation values $\phi_i$ and $\phi_i'$ of the field $\phi$ in the two bubbles, defined as 
\begin{equation}
    \cos(\theta) \equiv \frac{\phi_i\phi_i'}
    {|\phi||\phi'|} \,.
\end{equation}
We can therefore write, 
\begin{equation} \label{gravitywave}
    h_{ab} = \, |C|^2\, \int_{\theta} h_{ab}^{(\theta)} \,.
\end{equation}  
It is clear that the above expression is \textit{macroscopically} different from the waves sourced by classical mergers with particular values of $\theta$. 

As a specific example, let us consider the simplified case $N=1$, corresponding to a spontaneously broken $Z_2$-symmetry.  
Classically, there exist two types of bubble solutions $\phi(x)$, which differ by their sign.  
As already discussed, in the quantum  theory they are described by the coherent states \eqref{coh}, $\ket{{\rm bubble}}_{\pm}$, which satisfy $\bra{{\rm bubble}}_{\pm} \hat{\phi}(x) \ket{{\rm bubble}}_{\pm} = \pm \phi(x)$. 
The two represent the states that are $Z_2$-conjugated relative to each other. 
The $Z_2$-even and $Z_2$-odd bubble states are formed by the superpositions  
\begin{equation} \label{PMZ2_1}
    \ket{\pm} =  \frac{1}{\sqrt{2}} (\ket{{\rm bubble}}_{+}  \pm   \ket{{\rm bubble}}_{-})\, . 
\end{equation} 
Obviously, the state $\ket{+}$ is the same as the  $Z_2$-invariant state in \eqref{Z2} discussed earlier. 

Since prior to merger, the bubbles do not talk to each other, the state of the two bubbles is well approximated as a product state, 
\begin{equation} 
    \ket{\rm in} =  \ket{+} \otimes \ket{+}' \,.
\end{equation}
As previously discussed, we evolve this state in time, by acting with the time-evolution operator, ${\rm e}^{-i\hat{H}t}$.  
The resulting state can be split into a superposition of the classical states
\begin{equation} \label{tevolved}
    \ket{t} = \, \frac{1}{2} \, \sum_{A,B=\pm} \ket{\rm{cl}(t)}_{A,B} \,, 
\end{equation}
where 
\begin{equation} \label{CLtime}
    \ket{\rm{cl}(t)}_{A,B} \equiv {\rm e}^{-i\hat{H}t}\ket{{\rm bubble}}_{A}\otimes \ket{{\rm bubble}}_{B}'   
\end{equation}
represent the time evolved states, originating from different versions of the classical bubble mergers. 
After the mergers, these states are no longer tensor product states, but are still classical to a good approximation.
As already discussed, in this analysis we shall ignore the possible effects of quantum breaking. 

This purifies the effect we are after since, even under the robustness of the classical evolution, due to the microstate degeneracy, the bubble mergers cannot be described classically. 
Rather, the process is described by the quantum superposition of several classical trajectories.  
This phenomenon is independent of previously-discussed quantum breaking effects.

We thus see, that the time evolution of merging entangled bubbles, described by the quantum state $\ket{t}$, represents an entangled superposition of four classical trajectories $\ket{\rm{cl}(t)}_{A,B}$, corresponding to different classical mergers.

In a would-be classical evolution, the outcomes from these mergers are macroscopically different. 
For example, the classical trajectories $\ket{\rm{cl}(t)}_{+,+}$ and $\ket{\rm{cl}(t)}_{-,-}$ describe mergers in which the bubble walls annihilate and one ends up with a larger expanding bubble, uniformly filled with the $\phi = f$ or $\phi = -f$ vacuum, respectively. 

In contrast, the classical trajectories $\ket{\rm{cl}(t)}_{\mp}$ and $\ket{\rm{cl}(t)}_{\pm}$ will end up with a classical bubble, in the interior of which the two classical vacua are separated by a domain wall. 
The existence of the wall follows from topological considerations, since the two classical vacua $\phi = \pm f$ are exactly degenerate in energy. 

\begin{figure}
    \centering
    \begin{subfigure}{\linewidth}
        \includegraphics[width=\linewidth]{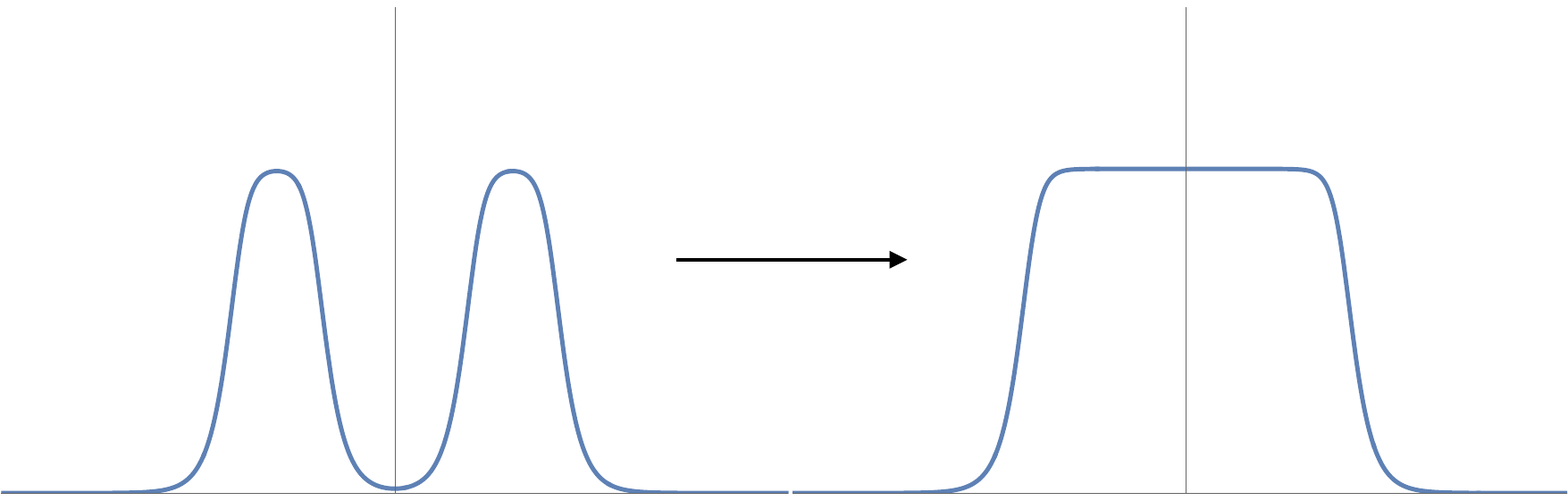}
        \label{fig:2bubbles}
        \caption{The merger of two bubble solutions, with the same sign.}
    \end{subfigure}
    \begin{subfigure}{\linewidth}
        \includegraphics[width=\linewidth]{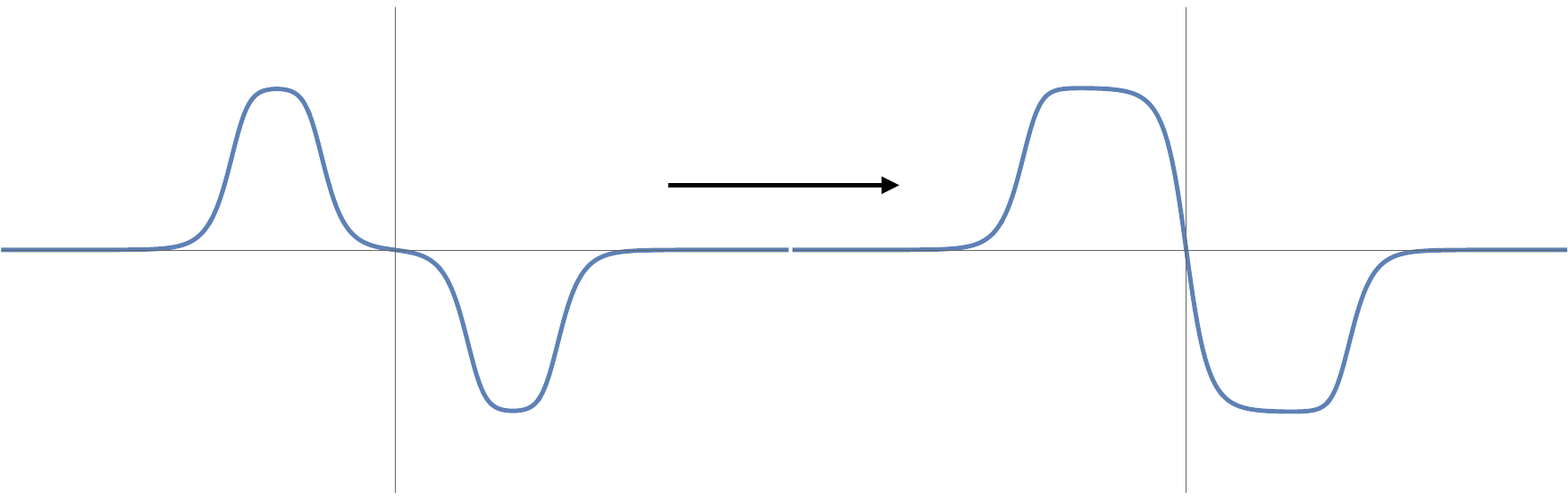}
        \label{fig:bubbleantibubble}
        \caption{The merger of two bubble solutions, with opposing signs.}
    \end{subfigure}    
    \caption{Illustration of the merger of two bubble solutions with the same signs, Fig. a), and of two bubbles merging, with opposite signs, Fig. b). }
        \label{fig:mergers}
\end{figure}

\begin{figure}
    \centering
    \begin{subfigure}{\linewidth}
        \includegraphics[width=0.7\linewidth]{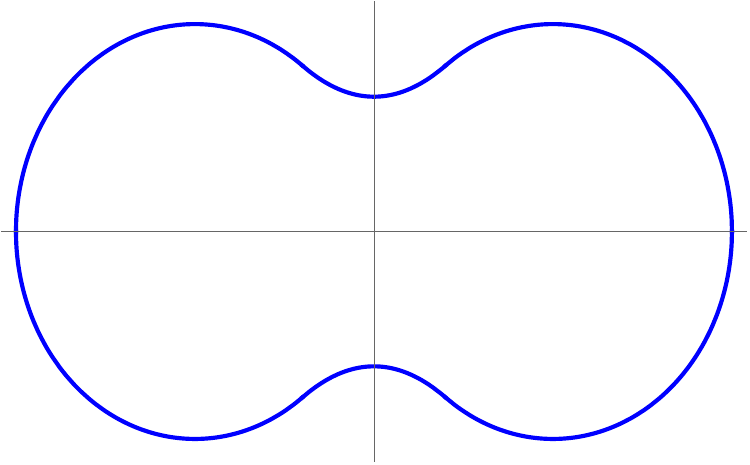}
        \label{fig:++bubble}
        \caption{The shape that forms after the merging of two bubbles where the interior vacuum is the same.}
    \end{subfigure}
    \begin{subfigure}{\linewidth}
        \includegraphics[width=0.7\linewidth]{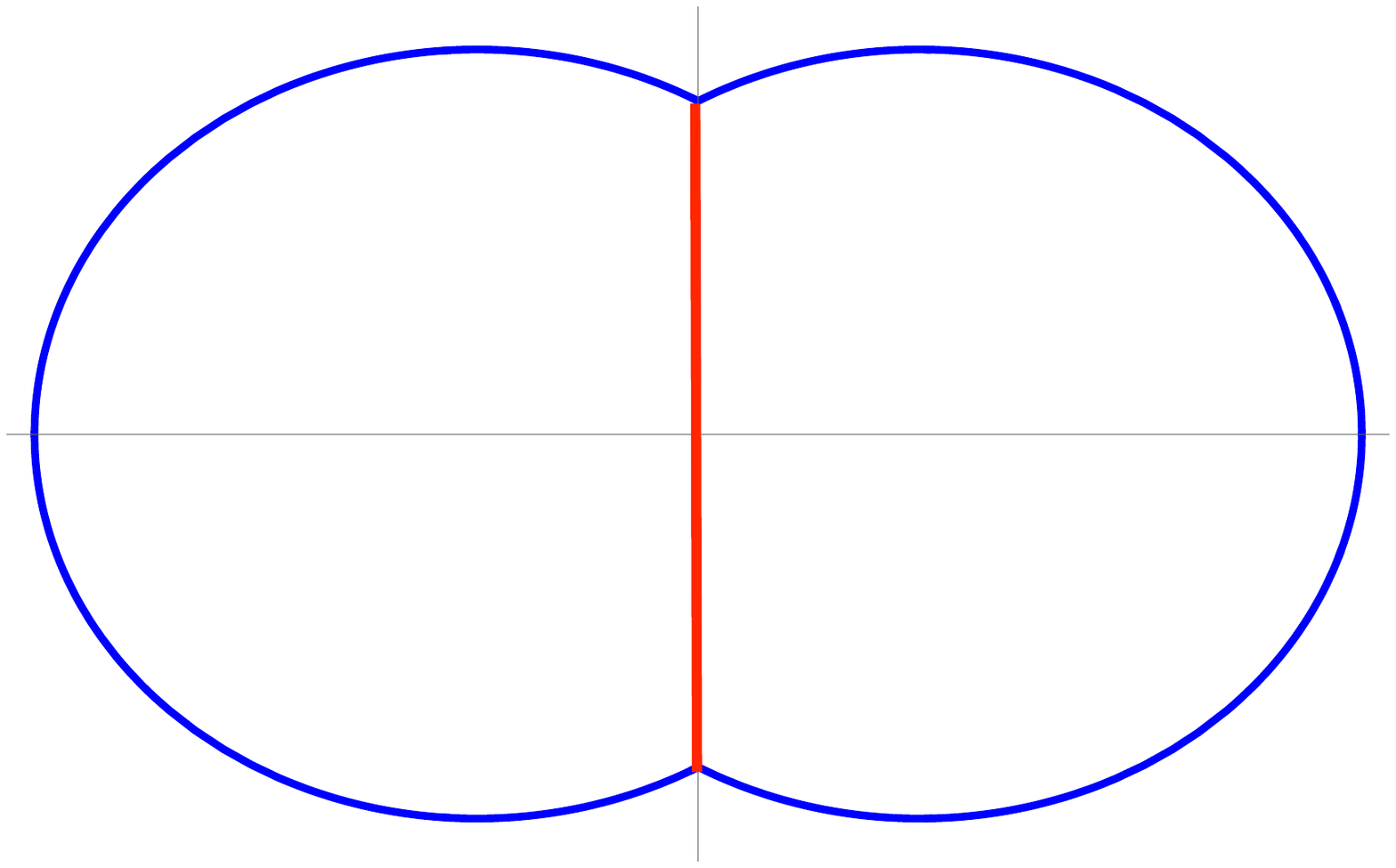}
        \label{fig:+-bubble}
        \caption{The shape that forms after the merging of two bubbles where the interior vacua have opposite signs. 
        The merging region is less smooth and forms a junction with the interior domain wall.}
    \end{subfigure}
    \caption{An Illustration of the shape after the merging of two vacuum bubbles. a) Illustrates the merged shape of two bubbles, where the vacuum in the interior is the same, whereas in b) the two interior vacua have opposing signs. 
    Therefore, the merging region contains a junction in the middle, where the bubble walls and the interior domain wall are joined. }
    \label{fig:bubbleshapes}
\end{figure}

Fig. \ref{fig:mergers} and Fig. \ref{fig:bubbleshapes} illustrate the two possible bubble mergers. 
In Fig. \ref{fig:mergers}, the merger is illustrated in terms of the profile functions, showing how the profile of two bubble solutions of the same sign simply create a larger bubble solution (Fig. \ref{fig:mergers} a)). 
In contrast, in the case of opposing signs, a domain wall forms in the middle (Fig. \ref{fig:mergers} b)).
Fig. \ref{fig:bubbleshapes} shows the shapes of the bubbles right after the mergers.
If the interior vacua of the bubbles are the same, the bubbles smoothly merge to become one (Fig. \ref{fig:bubbleshapes} a). Whereas, in the case of differing vacua, the two bubbles, 
being glued together, keep being separated by a domain wall, which forms a junction with the exterior bubble walls (Fig. \ref{fig:bubbleshapes} b)).

As already discussed, in the thin-wall regime, this wall can be thought of as being a composite of two "elementary" walls, separating the $\phi = \pm f$ vacua in the bubble. 
Indeed, at $M(T) =0$, in \eqref{V1} all three vacua become exactly degenerate, and the elementary walls repel each other. 
However, for $M(T) \neq 0$ the degeneracy is lifted, so that the vacua $\phi = \pm f$ become energetically more favourable. 
Correspondingly, the pressure difference confines the elementary walls, forming a composite wall, separating the $\phi = \pm f$ vacua. 

  
Now, the true quantum trajectory $\ket{t}$ represents a superposition of all the above classical trajectories. 
Obviously, the physical observables obtained from this superposition will be macroscopically different from the ones corresponding to each of the four trajectories.  
First, the measurements with external probes, sensitive to $Z_2$ quantum numbers, will detect statistically different outcomes. 
Secondly, the difference between quantum and classical mergers will be imprinted in gravitational radiation, even though the coupling of the graviton is $Z_2$-even. 

Indeed, considering the equation \eqref{gravity}, it is clear that the expectation value, taken over the state $\ket{t}$, given in \eqref{tevolved}, is different from the analogous expectation values taken over any of the four classical states. 
This is already clear from the understanding of classical mergers, which are very different in four differing cases. 

The expectation value $\bra{t} \hat{T}_{ab} \ket{t}$ represents the sum of classical energy momentum tensors describing each of the four classical mergers to a good approximation: 
\begin{equation}
    \bra{t} \hat{T}_{ab} \ket{t} = \frac{1}{2} \sum_{A,B} T_{ab}^{(A,B)}
\end{equation}
where $T_{ab}^{(A,B)} = \bra{\rm{cl}(t)}_{A,B}\hat{T}_{ab}\ket{\rm{cl}(t)}_{A,B}$.
Due to classicality of the states $\ket{\rm{cl}(t)}_{A,B}$, we ignore the off-diagonal 
matrix elements $\bra{\rm{cl}(t)}_{A,B}\hat{T}_{ab}\ket{\rm{cl}(t)}_{A',B'}$, with $A\neq A'$ and $B \neq B'$. 

The resulting \textit{classical} gravitational waves represent a linear superposition of gravitational waves emitted by four different classical mergers, 
\begin{equation}
    h_{ab} = h_{ab}^{(++)} + h_{ab}^{(--)}+  h_{ab}^{(+-)}+ h_{ab}^{(-+)} \,.
\end{equation}  
Due to the $Z_2$ invariance, with all other parameters fixed, we have $h_{ab}^{(++)} = h_{ab}^{(--)}$ and $h_{ab}^{(+-)}= h_{ab}^{(-+)}$. 
However, $h_{ab}^{(++)}$ and $h_{ab}^{(+-)}$ are \textit{macroscopically} different, due to the significantly different mergers. 
Thus, the gravitational wave detector, in the case of waves emitted by a true quantum merger, should see waves macroscopically different from any of the four possible classical mergers. 
   
\section{Quantum Breaking Effects from Particle Creation} 

Let us now take into account the effects of quantum breaking~\cite{Dvali:2013vxa, Dvali:2013eja, Dvali:2017eba, Dvali:2017ruz, Berezhiani:2021zst, Berezhiani:2020pbv, Berezhiani:2021gph, Berezhiani:2023uwt, Michel:2023ydf, Berezhiani:2025tkp}. 
As already discussed, the essence of the phenomenon is a departure of the true quantum evolution of the system from its classical trajectory. 
  
For an isolated, expanding bubble evolving from an initial classical state $\ket{{\rm lc}(0)}_A$, the main source of the departure is expected to come as the back-reaction from particle creation.  

The particle creation by an expanding classical bubble has been previously studied in the semiclassical regime~\cite{Aoyama:1982gg, Rubakov:1984pa, Yamamoto:1994te}. 
In this regime, the back reaction on the quantum state of the bubble is not taken into account.  
However, this back reaction is very important, as it affects the quantum state of the bubble. 
The effect is especially relevant for bubbles with a high microstate entropy. 

In the full quantum picture, the source of quantum breaking can be understood in the following way. 
Let us consider the time evolution of an initially classical bubble. 
In the quantum language, this bubble state represents a coherent state of the type \eqref{cohJ}~\cite{Saturon2020}.

On this state, the one-point faction of the quantum field $\hat{\phi}_j$ matches the classical bubble solution with a definite symmetry breaking pattern, $O(N) \rightarrow O(N-1)$. 
In other words, the state of the classical bubble is $O(N-1)$-invariant. 
At the same time, on a classical bubble state, the higher-point functions are negligible (or rather, factorizable in one-point functions). 

When bubble evolves in time, the constituents re-scatter. 
This re-scattering leads to the change of the bubble state and, in particular, to particle creation. 
As a result, the bubble state will ``diversify" in the $O(N)$-space. 
Of course, the total state of bubble plus the created radiation must remain $O(N-1)$-invariant. 
However, the two become highly entangled. 

Schematically, the evolution process of the state, initially describing a classical bubble with particular symmetry breaking pattern, can be represented as, 
\begin{equation}  \label{Bdecay}
\ket{{\rm cl}(t)}_A \rightarrow  \sum_{\eta} \ket{A}_{\eta}\otimes \ket{\rm rad}_{A,\eta} \,, 
\end{equation}
where the states $\ket{A}_{\eta}$ and  $\ket{\rm rad}_{A,\eta}$ denote various bubble and radiation parts of the evolved state
and the index $\eta$ stands for the set of all characteristics over which the entanglement takes place.
The label $A$ identifies the initial classical state. Neither of the components ($\ket{A}_{\eta}$ and  $\ket{\rm rad}_{A,\eta}$) are in general $O(N-1)$-singlets, only the total state is.  

In other words, the time-evolved state of an $O(N-1)$-invariant (initially) classical bubble $\ket{\rm cl(0)}_A$ represents an $O(N-1)$-invariant state of entangled $O(N-1)$-non-invariant bubble states and non-invariant radiation states.
 
Now, in the time evolution of the initially $O(N)$-invariant bubble state \eqref{QbubbleT}, each classical entry $\ket{\rm cl(t)}_A$ is subjected to the above quantum breaking effect. 

In particular, quantum breaking will manifest itself in the fact that the higher-point functions on the states $\ket{{\rm cl}(t)}_A$ no longer factorize into products of one-point functions. 
Of course, quantum breaking affects each classical bubble state as well as the would-be classical states $\ket{{\rm lc}(t)}_{A,B}$ formed in their mergers. 
  
As a result, the off-diagonal matrix elements $\bra{\rm{cl}(t)}_{A,B}\hat{T}_{ab}\ket{\rm{cl}(t)}_{A',B'}$ will no longer vanish. 
Correspondingly, the expectation value of $\hat{T}_{ab}$ in \eqref{Tvev}  can no longer be represented as the sum of the classical entries. 
Consequently, the gravitational waves emitted from such a merger cannot be reduced to a superposition of waves sourced by the classical mergers as in eq. \eqref{waveAB}. 

As an illustration, let us consider quantum breaking effects for an expanding  $Z_2$-invariant bubble \eqref{Z2}.  
The emitted particles form a state which we shall call radiation. 
We denote by  $\ket{\rm rad}_{\pm}$ the radiation states that are $Z_2$-even and odd respectively.

For illustrative purposes, using the bubble and radiation states, we can form a one-parameter family of $Z_2$-invariant (i.e., $Z_2$-even) states.
\begin{equation} \label{PMZ2_2}
    \ket{\eta} =  \cos(\eta) \ket{+} + \sin(\eta) \sum_{\pm} 
    \ket{\pm}_r\times \ket{\rm rad}_{\pm} \,,
\end{equation} 
where $\eta$ is a parameter and $\ket{\pm}_r$ stand for the back-reacted
$Z_2$-even and $Z_2$-odd bubble states.
The first entry describes a $Z_2$-invariant bubble without radiation back reaction.
Of course, more complicated states can readily be constructed, but the above simplified version suffices to make the point. 

Let us imagine a bubble that time evolves along a trajectory parameterized by $\eta$, with $\eta=0$ the initial state.  
As $\sin(\eta)$ increases, the bubble degrees of freedom become more and more entangled with radiation.
On such a state the non-factorizable higher-point functions grow and can eventually compete with the one-point one. 

The influence of higher-point functions will be imprinted in classical gravitational radiation, since the  expectation value \eqref{Tvev} is not reducible to the sum of would-be classical entries. 

\section{Quantum chaos}

The first order phase transitions in which bubbles have high microstate degeneracy can be assigned a well-defined measure of chaoticity in the sense of the paper~\cite{Dvali:2013vxa}. 
The physical meaning of chaoticity is the sensitivity of the  final state with respect to initial perturbations. 

The paper~\cite{Dvali:2013vxa} has investigated an unstable system which can evolve in a large number of degenerate trajectories. 
The choice of the trajectories is highly sensitive to small initial perturbations.  
Correspondingly, weak initial perturbations make the system evolve into macroscopically different final states.   

In the example studied in~\cite{Dvali:2013vxa}, the unstable state possesses a Lyapunov exponent, implying that it is unstable already at a classical level. 
In the first order phase transition discussed in the present paper, the  false "vacuum" is classically stable and the instability manifests itself at the level of quantum tunneling. 
However, to define chaoticity, this difference is inessential.  

In the present case too, the state of the materialized bubble as well as its subsequent evolution are highly sensitive to initial perturbations. 

It is important to understand that chaoticity is not a property of a particular state or a trajectory, but rather a property of the Hamiltonian to possess a high degeneracy of diverse trajectories. 

It is clear that this concept of chaoticity is directly applicable to the present case. 
The $O(N)$-invariant unstable "vacuum" can tunnel only into an $O(N)$-symmetric bubble described by eq. \eqref{invB}, which represents a superposition of an exponentially large number of microstate bubbles, given by eq. \eqref{sumMCR}. 
However, this state is very sensitive to initial perturbations.  
For example, an initial perturbation that preserves a subgroup of $O(N)$, can affect the materialized bubble macroscopically.  

We can quantify the above effect by the chaoticity parameter, which we define as the ratio of the bubble entropy to the saturated value determined by the bound \eqref{Alpha}, 
\begin{equation} \label{Xi} 
    \xi \equiv \frac{S_{\rm bubble}}{S_{\rm max}} \,.
\end{equation} 
For a thick wall bubble this will take the form 
 \begin{equation} \label{XiTH} 
    \xi \equiv \alpha \, S_{\rm bubble} \,. 
\end{equation} 
  
Thus, chaoticity is an intrinsic feature of phase transitions with a high degeneracy of bubble microstates. 
It is a direct consequence of the sharp increase of the number of available microstates after the transition.   
The density of states in the bubble state as compared to the initial $O(N)$-invariant false vacuum is exponentially large.  
Correspondingly, so is the density of Euclidean trajectories leading to the materialization of the bubble. 

During the transition, the system explores all trajectories simultaneously and correspondingly ends up in a highly entangled  $O(N)$-invariant superposition \eqref{invB} of microstates.    
Due to this, small changes in the initial state can lead to large variations in the final state. 
In other words, the bubble acquires a memory load from the initial $O(N)$-violating perturbation.  
The initial perturbation projects the bubble to a subspace of the memory space. 
Further diversification of trajectories takes place during the subsequent evolution of the bubble, as discussed above. 
 
In different bubbles the residual memory loads are uncorrelated and can therefore differ significantly. 
This misalignment of the entangled information patterns results in macroscopic inhomogeneities that source gravitational waves and get imprinted in their spectrum.  

\section{Comparing the Swift Memory Burden Effects in Bubbles versus Black Holes} 
 
The influence of the internal quantum microstate on the classical dynamics of the bubble merger can be viewed as a special case of the \textit{swift memory burden effect} \cite{Dvali:2025sog}. 
This effect is generically expected to take place in mergers of black holes as well as in those of solitons, and can significantly affect the dynamics of the process. 
In particular, it influences the spectroscopy of the gravitational radiation from mergers.  

The essence of the phenomenon is that the quantum microstate of the object is described by the excitation patterns of internal gapless modes (memory modes).
The features of the excitation pattern are clasically invisible in the ground state.
However, they strongly influence the dynamics of the system upon its perturbation. 
Naturally, the effect is more prominent in saturons due to their maximal microstate degeneracy which saturates the bounds \eqref{Alpha} and \eqref{Area}.
However, it can have a significant effect even if the microstate degeneracy is well below these bounds. 

Due to the above similarities, it is instructive to compare vacuum bubbles and black holes in more detail.  
A comparative study of the memory burden effect in black holes and in solitons was previously given in \cite{Dvali:2024hsb}.  
Specifically, the swift memory burden effect in these two systems was discussed in \cite{Dvali:2025sog}.
The previous two papers shall guide the following discussion, with the soliton part specifically adjusted to vacuum bubbles.  

Before discussing similarities and differences between the swift memory burden effects in mergers of vacuum bubbles versus the mergers of black holes, we shall first compare some features of these two entities when they are in isolation.   
   
\subsection{Microstate Entropies} 

In both cases (black holes versus vacuum bubbles) the information is encoded in the set of memory modes, which are gapless in the interior of the object. 
That is, both systems operate via the mechanism of \textit{assisted gaplessness} which was originally identified in~\cite{Dvali:2017nis, Dvali:2017ktv, Dvali:2018vvx, Dvali:2018xoc, Dvali:2018tqi} as the universal mechanism of efficient information storage and was shown to lead to memory burden effect~\cite{Dvali:2018xpy, Dvali:2020wft}. 
Correspondingly, for unperturbed objects all information patterns are degenerate in energy. 
This feature is shared by black holes and vacuum bubbles. 
   
As already discussed in a series of papers \cite{ S1,S2,Saturon2020,S3,S4,S5,S6, Dvali:2024hsb, Contri:2025eod}, similar to a black hole, an isolated saturated soliton or a vacuum bubble with spontaneous symmetry breaking carries a microstate entropy given by the expressions \eqref{AlphaB} and  \eqref{AreaB}. 
This is exactly the same expression as is satisfied by the Bekenstein-Hawking entropy, with the substitution $f=M_P$. 
This is not an accident, since, as explained previously in the above papers, the scale at which Poincar\'e symmetry breaking occurs in a black hole, is given by $M_P$. 
  
As was discussed in~\cite{SaturonDM} and in the present work, this feature is fully inherited by expanding vacuum bubbles. 
The difference is that the bubble falls out of the saturation regime, due to the expansion. 
Of course, as the radius grows, the absolute value of the entropy grows as well, but not as fast as the area of the bubble.  
Correspondingly, for a thin-wall bubble with $R \gg m^{-1}$, the  entropy becomes smaller than the saturated value \eqref{Area}. 
 However, this feature does not introduce a qualitative difference in the present discussion. 
 First, regardless of saturation, a  bubble is materialized 
 in a maximally entangled state \eqref{invB}. Second, during the bubble expansion the entanglement only grows, involving the created particles \eqref{Bdecay}. 
Correspondingly, in the moment of the bubble merger the entanglement is maximal.

\subsection{Information Horizons of Bubbles and Black Holes} 
   
In the semi-classical description, a black hole possesses an information horizon.  
As discussed previously \cite{ S1,S2,Saturon2020,S3,S4,S5,S6, Dvali:2024hsb, Contri:2025eod}, the same feature is exhibited by all saturons and, in particular, by saturated vacuum bubbles  \cite{Saturon2020,S4, Dvali:2024hsb, Contri:2025eod}. 
The bubble horizon has a very transparent physical meaning.

Indeed, the information-carrying memory modes are the gapless Goldstone bosons of the spontaneously broken $O(N)$-symmetry. 
The information is encoded in their $O(N)$ quantum numbers. 
Correspondingly, the retrieval of information requires the detection of the $O(N)$-flavor state of the gapless Goldstone modes. 
However, outside of the bubble, the theory possesses a mass gap $\sim m$.  
Correspondingly, a {\it passive} extraction of information, via particle emission, is not possible for a stationary bubble and is highly suppressed for an expanding one. 

On the other hand, an {\it active}  extraction of information, via the scattering of some probe particles at the bubble, takes time 
~\cite{Saturon2020,Dvali:2021jto, S3, S4, Dvali:2024hsb, Contri:2025eod}
\begin{equation} \label{Rtime}
 t_{\rm inf} \sim \frac{R}{\alpha} \sim S\, R \,.  
\end{equation}  
In the semi-classical limit of the theory, this time goes to infinity. 

Correspondingly, a semi-classical bubble, very similarly to a semiclassical black hole, possesses an exact information horizon. 
    
Of course, for a black hole, the horizon is universal. 
Indeed, since gravity interacts with all sources, there exist no particles for which the black hole can be transparent. 
In contrast, the horizon of the vacuum bubble only applies to probes with $O(N)$-quantum numbers.  
However, this difference is trivial. 
The important feature is that each object possesses a horizon for all probes experiencing the interaction that forms the respective object.  

\subsection{No-hair features of bubbles and black holes}
    
A classical black hole in its ground state obeys the so-called \textit{no-hair} theorems~\cite{Ruffini:1971bza, Bekenstein:1972ny, Bekenstein:1971hc, Bekenstein:1972ky, Teitelboim:1972pk, Teitelboim:1972ps, Hartle:1971qq}.
These theorems state that a classical black hole is fully characterized by only a few quantum numbers. 
These include the mass, the angular momentum and electric and/or magnetic charges with respect to long-range gauge interactions. 
All the additional information, no matter how rich, is completely hidden for an outside observer. 
 
Interestingly, in the semi-classical regime, all saturons, including vacuum bubbles, have similar no-hair properties \cite{S1,S2, Saturon2020, Dvali:2021jto,S3,S4,S5,S6, Dvali:2024hsb, Contri:2025eod}. 
These features follow from the time-scales required to retrieve the information, as discussed above. 
    
For an observer, living in an $O(N)$-invariant vacuum outside a classical bubble, the information content stored inside the bubble is unreachable. 
This feature is shared equally by stationary and expanding bubbles. 
Of course, expanding bubbles, by default, have no ground state.
However, the above no-hair feature is consistently defined for a distant observer up until it encounters the bubble. 
 
Notice that, even after the observer gets ``swallowed" by the bubble, the read-out of the information stored in the bubble interior can only begin after the time  \eqref{Rtime}~\cite{Saturon2020,Dvali:2021jto, S3, S4, Dvali:2024hsb, Contri:2025eod}.
   
\subsection{Swift Memory Burden in Bubbles versus Black Holes} 
   
\subsubsection{Black holes} 
   
The swift memory burden phenomenon~\cite{Dvali:2025sog} tells us that a black hole of mass $M$ is characterized by an additional quantity, the memory burden parameter $\mu$,
\begin{equation} \label{MUBH}
    \mu \equiv \frac{M}{E_p},
\end{equation} 
where, $E_p$ is the would-be energy cost of the information pattern in the absence of a black hole. 
As explained in~\cite{Dvali:2025sog}, the quantity $\mu$ is not in conflict with the black hole no-hair theorems. 
First,  it is quantum in origin. 
Most importantly, as long as the black hole is in its ground state, $\mu$ has no effect.
However, memory burden gets activated swiftly for a perturbed black hole. 
Upon perturbation, the memory load exerts a backreaction on the classical dynamics.
In particular, the spectroscopy of black hole mergers is highly sensitive to the parameter $\mu$.

At the microscopic level, the story is the following. 
For a black hole, the memory modes come from zero energy excitations of various angular momenta \cite{Dvali:2011aa, Dvali:2012en}. 
Around the state of an unperturbed classical black hole, these modes are all gapless. 
Correspondingly, in this macroscopic state, all the microstates described by distinct information patterns are degenerate. 
However, when the black hole is perturbed classically, two things happen: 
  
First, the information patterns become costly in energy. 
Second, the patterns split in energy and the degeneracy lifts. 
When a black hole, with radius $R$, is in its ground state, an exponentially large number of patterns populate a tiny energy gap $\Delta E \sim 1/R$. 
For a classically perturbed black hole with maximal memory load, the energy spread of the patterns becomes comparable to the mass of the black hole.

\subsubsection{Bubbles} 
   
Similarly to black holes, the internal quantum state of a bubble affects the merger process of two bubbles.  
The role of the memory modes of the bubble is taken up by the Nambu-Goldstone modes, which determine its internal quantum state.  
Therefore, one can define the memory burden parameter that quantifies the fraction of activated memory modes.
 
For a non-expanding bubble, discussed in \cite{Saturon2020,S4, Dvali:2024hsb, Contri:2025eod}, the $\mu$ parameter can be defined via \eqref{MUBH}, exactly as for a black hole.  
   
However, for expanding vacuum bubbles, the expression \eqref{MUBH} is not a good measure for the swift memory burden effect.  
Since a bubble created via tunneling can have zero energy, a better characteristic is the chaoticity parameter $\xi$, rather than $\mu$. 
As discussed, this parameter measures the available memory space as compared to the maximal value given by the bounds \eqref{Alpha} and \eqref{Area}. 
       
Now, for a vacuum bubble, the memory modes are the Nambu-Goldstone modes of the spontaneously broken $O(N)$ symmetry. 
Correspondingly, the patterns remain degenerate in energy as long as the bubble keeps expanding in the vacuum with an unbroken $O(N)$ symmetry. 
External perturbations and particle creation can affect the symmetry properties of the bubble pattern, however there are still many  patterns that remain degenerate under the various subgroups of $O(N)$. 
                 
Upon an encounter with another bubble, with a misaligned memory pattern, memory burden sets in. 
This is similar to the swift memory burden effect in mergers of solitons with misaligned memory patterns \cite{Dvali:2025sog}.  
     
Another difference between black holes and vacuum bubbles lies in the entanglement of the memory patterns.   
Unlike a black hole, all vacuum bubbles, created as a result of tunneling from the $O(N)$-invariant vacuum, materialize in the same $O(N)$ invariant state \eqref{invB}. 
In this state, all the memory patterns are maximally entangled. 
In \cite{Dvali:2025sog}, the information carried by the entangled information patterns was said to be \textit{shuffled-up}.   
Thus, one can say that in the $O(N)$-invariant bubble, the information is shuffled-up
maximally. 
   
The misalignment of patters between the two bubbles takes place due to the subsequent time evolution. 
As already discussed, particle creation and external perturbations affect the $O(N)$-invariance of the bubble state.  
In particular, each evolving bubble becomes entangled with outgoing radiation, with each member of the superposition being in a different $O(N)$-state.  

As shown in \cite{Dvali:2025sog}, the story with black holes is different. 
First, the memory patterns of two black holes already differ in their respective moment of formation. 
This is because black holes are not produced as a result of tunneling from a featureless vacuum, but rather from the collapse of highly energetic objects. 
The collapsing object (e.g., a star) usually carries a rich diversity of features.
After the collapse, these features get encoded in the black hole information patterns.  
   
Also, unlike what happens in vacuum bubbles, in a black hole the different information patterns may or may not be significantly entangled. 
In other words, the black hole information does not have to be shuffled up.  This depends on the diversity of features 
of the source that created it.  
In this sense, swift memory burden in black holes takes place regardless of the level of the entanglement.  
For merging black holes, swift memory burden is relevant as long as the total information load is not zero.  

\section{Conclusions and Outlook} 

The microstate degeneracy can introduce qualitatively new features in phase transitions. 
In particular, it affects the state and the subsequent time evolution of the nucleated vacuum bubble. 
No matter how macroscopic, the bubble state is not describable classically. 
Instead, by symmetry, the bubble is materialized in a maximally entangled quantum superposition of the degenerate would-be classical bubbles.

The microstate degeneracy also enhances the transition rate. 
The maximal enhancement is reached in the saturon regime which takes place when the bubble entropy saturates the QFT upper bounds \eqref{Alpha} and \eqref{Area}. 
As already discussed in \cite{SaturonDM}, in this limit the microstate entropy of the bubble state compensates the exponential suppression of the nucleation rate.
Correspondingly, the theory enters a qualitatively new regime, referred to as the \textit{saturon transition}. 
As we have seen, simultaneously, the bounds \eqref{Alpha} and \eqref{Area} are saturated by the corresponding Euclidean bounce.

Saturation of unitarity by the transition amplitudes and the related regime change is generic for the scattering processes in which the final state saturates the bounds \eqref{Alpha} and \eqref{Area}~\cite{Saturon2020}.  

In particular, the saturation phenomenon in bubble nucleation is closely linked with the observation made in~\cite{S2}, that in Yang-Mills theory the gauge instanton saturates the bounds \eqref{Alpha} and \eqref{Area} when the theory approaches the confining regime. 
Both cases represent manifestations of a general effect connecting the regime-change of the vacuum transitions with the saturation of the bounds \eqref{Alpha} and \eqref{Area} by the Euclidean trajectories. 
 
However, more importantly, the microstate degeneracy introduces new features even when it is well below the saturation points of \eqref{Alpha} and \eqref{Area}. 
Namely, each bubble materializes in a maximally entangled quantum superposition \eqref{invB} of microstates. 
Due to this, by the time the bubbles overlap and collide, they are fully quantum and highly entangled. 

This produces a qualitatively new source of gravitational waves, which cannot be reduced to any particular classical merger. 
The resulting gravitational wave spectrum can be fundamentally different from the ones sourced by previously considered classical inhomogeneities, such as (non-entangled) bubbles from first order transitions~\cite{ Kosowsky:1991ua, *Kosowsky:1992rz, FirstOrder1}, collapsing defects~\cite{Vilenkin:1981bx, Vachaspati:1984yi, Martin:1996ea, Martin:1996cp, Gleiser:1998na} (see also, some recent updates on gravitational waves from cosmic string loops~\cite{Wachter:2024zly} or confined monopoles~\cite{Dvali:2022vwh}), and various forms of turbulence~\cite{Turbulence1, Turbulence2, Turbulence3}. 

Of course, any other type of radiation, potentially sourced by the mergers of entangled bubbles, shall be affected similarly.   

The phase transitions with highly degenerate vacuum bubbles are characterized by quantum chaos defined in the sense of 
~\cite{Dvali:2013vxa}. The chaoticity of the system can be measured by the parameter $\xi$, given in \eqref{Xi}. 
This quantity represents the ratio of the bubble microstate entropy to the bound \eqref{Alpha}. 
The maximal value, $\xi \sim 1$, is reached in saturon phase transitions. 
   
The discussed influence of the microstate degeneracy on the coalescence of two bubbles shares a similarity with the \textit{swift memory burden effect} in merging black holes \cite{Dvali:2025sog}.
In both cases, the classical dynamics of the merger is affected by the quantum state of the internal degrees of freedom (the memory modes) of the merging objects. 

For an isolated, unperturbed object, the internal microstate is unresolvable classically.
However, it strongly influences the process of the merger. 
The effect is macroscopic and gets imprinted in the resulting gravitational radiation. 
Thus, mergers with identical classical initial conditions can have macroscopic variations in the gravitational wave spectrum, depending on the internal quantum state.  
The parallels and differences in the cases of bubbles versus black holes were discussed in detail. 
 
In order to better quantify the swift memory burden effect and the chaos in merging vacuum bubbles, it is necessary to evaluate the quantum dynamics of an expanding bubble more precisely. 
The right approach to this is to view the classical vacuum bubble as a coherent state of field quanta \cite{Saturon2020}. 
Then, a fully quantum time-evolution of the coherent state can be performed by considering particle creation by the classical bubble as a multi-particle process in the spirit of \cite{Dvali:2022vzz}.
This approach makes it possible to view non-perturbative semiclassical effects, such as particle creation or instanton transitions, as fully quantum multi-particle processes. 
Due to this, it is possible to capture the effects of quantum back-reaction.

Such a quantum resolution of a classical field was originally proposed for black holes \cite{Dvali:2013eja} and de Sitter \cite{Dvali:2013eja, Dvali:2017eba, Berezhiani:2021zst}, where both were described as coherent states of gravitons. 
This analysis, in the semiclassical limit, reproduces the Hawking and Gibbons-Hawking effects, respectively.  
An analogous, fully quantum analysis of particle-creation by various saturated solitons was given in \cite{S3,S4, Dvali:2024hsb, Contri:2025eod}.

Alternatively, the coherent state of the bubble can be time evolved via a background field method \cite{Berezhiani:2020pbv, Berezhiani:2021gph, Berezhiani:2023uwt, Berezhiani:2025tkp}.

Finally, as argued in \cite{CGC1}, the Standard Model contains a candidate saturon state in the form of the color glass condensate of gluons \cite{CGC2} (for a related analysis, see \cite{GarciaTello:2024lie, Kou:2025mvt}).  
If so, this could have resulted in a saturon phase transition at temperatures well above the QCD scale, at which the saturated gluon condensate can form.  \\

{\textsl{\bf Acknowledgments}}\;---\;

 We thank Lasha Berezhiani, Otari Sakhelashvili and Tongxuan Zhang for discussions. 
 
This work was supported in part by the Humboldt Foundation under the Humboldt Professorship Award, by the European Research Council Gravities Horizon Grant AO number: 850 173-6, by the Deutsche Forschungsgemeinschaft (DFG, German Research Foundation) under Germany's Excellence Strategy - EXC-2111 - 390814868, and Germany's Excellence Strategy under Excellence Cluster Origins. \\
 
Disclaimer: Funded by the European Union. 
Views and opinions expressed are however those of the authors only and do not necessarily reflect those of the European Union or European Research Council.
Neither the European Union nor the granting authority can be held responsible for them.

\section*{Appendix} 

 In this appendix, following \cite{Saturon2020, S4}, we shall give an example of a vacuum bubble with saturated microstate entropy in a fully renormalizable theory. 
 
 We consider a theory of a scalar field $\phi$ in the adjoint representation of $SU(N)$
 \footnote{Notice that a symmetric representation of $SO(N)$ would work the same 
 way.}.  
 We take  $\phi_{i}^{j}$ as an $N \times N$ traceless Hermitian matrix with $i, j= 1, \dotsc, N$ the 
 $SU(N)$-indices. The zero temperature Lagrangian density is
\begin{eqnarray}
\label{model_vac_bub}
\mathcal{L} & = & \dfrac{1}{2} \text{tr} \left[ \left( \partial_{\mu} \phi \right) \left( \partial^{\mu} \phi \right) \right] - V\left[ \phi \right] \,, \\ \nonumber
{\rm where}, \quad V\left[ \phi \right] &=& \dfrac{\alpha}{2} \text{tr} \left[ \left( f \phi - \phi^2 + \dfrac{I}{N} \text{tr} \left[ \phi^2 \right] \right)^2 \right] \, ,
\end{eqnarray}
where, $I$ is the unit $N \times N$ matrix, $\alpha$ is a dimensionless coupling constant and $f$ is the scale of symmetry breaking.  

Since the theory is renormalizable, the only QFT constraint on the validity of the description is \cite{Saturon2020}: 
\begin{equation} \label{Ubound1}
    \alpha \lesssim \dfrac{1}{N} \, .
\end{equation} 
The vacuum equations,
\begin{equation}
    f\phi^i_j -(\phi^2)^i_j +\dfrac{\delta^i_j}{N} \text{tr} \left[\phi^2 \right] = 0 \,,
\end{equation}
admit several solutions, corresponding to vacua with different unbroken symmetries. 
These include the vacuum with unbroken $SU(N)$ symmetry, $\phi=0$, and vacua with the spontaneous symmetry breaking patterns $SU(N) \rightarrow SU(N-K) \times SU(K) \times U(1)$, where $0<K<N$. 
By construction, at zero temperature, all of these vacua are degenerate in energy. 
For example, in the vacuum with $K = 1$  the VEV of the field is
\begin{equation} \label{nonzerocomponent}
    \phi_{i}^{j} = \dfrac{f}{N-2} \text{diag} \left( (N-1), -1, \dotsc, -1 \right) \, .
\end{equation}
In this vacuum, the unbroken symmetry group is $SU(N-1) \times U(1)$,  and the spectrum contains $2(N-1)$ gapless Goldstone bosons. 
In contrast, in the vacuum with unbroken $SU(N)$ symmetry, there exist no gapless excitations, as the theory exhibits a mass gap,
\begin{equation} \label{mass}
    m = \sqrt{\alpha} f \, .
\end{equation}

Now, due to the gapless Goldstone modes, a bubble of a broken symmetry vacuum, embedded in a symmetric one, is highly degenerate. 
The microstate counting, originally given in \cite{Saturon2020, S4},  is the same as in the example discussed in the main text (up to unimportant numerical factors).  

Since the thermal corrections lift the degeneracy of the free energy among different symmetry breaking patterns, vacuum bubbles with broken $SU(N)$-symmetry can be produced via tunneling from the symmetric vacuum  at  a non-zero temperature.
For example, similarly to the model discussed in the main text, a thermal mass of the field $\phi$ can be generated by a coupling of the type \eqref{PhiChi},
\begin{equation} \label{SingletAdj}
    - \,\beta \chi^2 \text{tr}(\phi^2) \,, 
\end{equation} 
with a singlet  field $\chi$, which forms a thermal bath of temperature $T$. 
Even if the $\phi$-field is not in thermal equilibrium with $\chi$, the coupling \eqref{SingletAdj} 
will modify the free energy of the $\phi$-field as in \eqref{V1}, giving
\begin{align}  \label{Vadj}
    V(\phi, T)\,  = &  - \frac{M(T)^2}{2} \text{tr} \phi^2  +  \nonumber \\ 
    + & \frac{\alpha}{2} \text{tr} \left[ \left( f \phi - \phi^2 + \dfrac{I}{N} \mathrm{tr} \left[ \phi^2 \right] \right)^2 \right]  \,.     
\end{align} 
The overall behavior of this theory, regarding phase transitions and bubble nucleation is qualitatively the same as in the main text, and will therefore not be repeated here.  
However, there are some minor modifications: 
 
First, the unitarity constraint on the coupling $\beta$ is changed from \eqref{BetaB} to 
\begin{equation} \label{UbetaApp}
    \beta N^2 \lesssim 1\,.
\end{equation}    
Secondly, because the number of species for the $SU(N)$-adjoint field scales as $\sim N^2$, according to the general relation of~ \cite{Dvali:2007hz, Dvali:2007wp}, 
the gravitational species scale is changed from \eqref{species} to  
\begin{equation} \label{speciesAdj} 
    \Lambda_{\rm gr} = \frac{M_P}{N} \,.
\end{equation} 
    
\bibliographystyle{utphys}
\bibliography{citations.bib}

@article{Saturon2020,
    author = "Dvali, Gia",
    title = "{Entropy Bound and Unitarity of Scattering Amplitudes}",
    eprint = "2003.05546",
    archivePrefix = "arXiv",
    primaryClass = "hep-th",
    doi = "10.1007/JHEP03(2021)126",
    journal = "JHEP",
    volume = "03",
    pages = "126",
    year = "2021"
}

@article{SaturonDM,
    author = "Dvali, Gia",
    title = "{Saturon Dark Matter}",
    eprint = "2302.08353",
    archivePrefix = "arXiv",
    primaryClass = "hep-ph",
    month = "2",
    year = "2023"
}

@article{S1,
    author = "Dvali, Gia",
    title = "{Area Law Saturation of Entropy Bound from Perturbative Unitarity in Renormalizable Theories}",
    eprint = "1906.03530",
    archivePrefix = "arXiv",
    primaryClass = "hep-th",
    doi = "10.1002/prop.202000090",
    journal = "Fortsch. Phys.",
    volume = "69",
    number = "1",
    pages = "2000090",
    year = "2021"
}

@article{S2,
    author = "Dvali, Gia",
    title = "{Unitarity Entropy Bound: Solitons and Instantons}",
    eprint = "1907.07332",
    archivePrefix = "arXiv",
    primaryClass = "hep-th",
    doi = "10.1002/prop.202000091",
    journal = "Fortsch. Phys.",
    volume = "69",
    number = "1",
    pages = "2000091",
    year = "2021"
}

@article{S3,
    author = "Dvali, Gia and Sakhelashvili, Otari",
    title = "{Black-hole-like saturons in Gross-Neveu}",
    eprint = "2111.03620",
    archivePrefix = "arXiv",
    primaryClass = "hep-th",
    doi = "10.1103/PhysRevD.105.065014",
    journal = "Phys. Rev. D",
    volume = "105",
    number = "6",
    pages = "065014",
    year = "2022"
}

@article{S4,
    author = "Dvali, Gia and Kaikov, Oleg and Berm\'udez, Juan Sebasti\'an Valbuena",
    title = "{How special are black holes? Correspondence with objects saturating unitarity bounds in generic theories}",
    eprint = "2112.00551",
    archivePrefix = "arXiv",
    primaryClass = "hep-th",
    doi = "10.1103/PhysRevD.105.056013",
    journal = "Phys. Rev. D",
    volume = "105",
    number = "5",
    pages = "056013",
    year = "2022"
}

@article{S5,
    author = {Dvali, Gia and K\"uhnel, Florian and Zantedeschi, Michael},
    title = "{Vortices in Black Holes}",
    eprint = "2112.08354",
    archivePrefix = "arXiv",
    primaryClass = "hep-th",
    doi = "10.1103/PhysRevLett.129.061302",
    journal = "Phys. Rev. Lett.",
    volume = "129",
    number = "6",
    pages = "061302",
    year = "2022"
}

@article{S6,
    author = {Dvali, Gia and Kaikov, Oleg and K\"uhnel, Florian and Valbuena-Bermudez, Juan Sebastian and Zantedeschi, Michael},
    title = "{Vortex Effects in Merging Black Holes and Saturons}",
    eprint = "2310.02288",
    archivePrefix = "arXiv",
    primaryClass = "hep-ph",
    doi = "10.1103/PhysRevLett.132.151402",
    journal = "Phys. Rev. Lett.",
    volume = "132",
    number = "15",
    pages = "151402",
    year = "2024"
}

@article{Dvali:2022vzz,
    author = "Dvali, Gia and Eisemann, Lukas",
    title = {Perturbative understanding of nonperturbative processes and quantumization versus classicalization},
    eprint = "2211.02618",
    archivePrefix = "arXiv",
    primaryClass = "hep-th",
    doi = "10.1103/PhysRevD.106.125019",
    journal = {Phys. Rev. D},
    volume = "106",
    number = "12",
    pages = "125019",
    year = {2022}
}

@article{Bekenstein:1973ur,
    author = "Bekenstein, Jacob D.",
    title = "{Black holes and entropy}",
    doi = "10.1103/PhysRevD.7.2333",
    journal = "Phys. Rev. D",
    volume = "7",
    pages = "2333--2346",
    year = "1973"
}

@article{FirstOrder1,
    author = "Kamionkowski, Marc and Kosowsky, Arthur and Turner, Michael S.",
    title = "{Gravitational radiation from first order phase transitions}",
    eprint = "astro-ph/9310044",
    archivePrefix = "arXiv",
    reportNumber = "IASSNS-HEP-93-44, FERMILAB-PUB-93-235-A",
    doi = "10.1103/PhysRevD.49.2837",
    journal = "Phys. Rev. D",
    volume = "49",
    pages = "2837--2851",
    year = "1994"
}

@article{Turbulence1,
    author = "Kosowsky, Arthur and Mack, Andrew and Kahniashvili, Tinatin",
    title = "{Gravitational radiation from cosmological turbulence}",
    eprint = "astro-ph/0111483",
    archivePrefix = "arXiv",
    reportNumber = "RAP-334",
    doi = "10.1103/PhysRevD.66.024030",
    journal = "Phys. Rev. D",
    volume = "66",
    pages = "024030",
    year = "2002"
}

@article{Turbulence2,
    author = "Gogoberidze, Grigol and Kahniashvili, Tina and Kosowsky, Arthur",
    title = "{The Spectrum of Gravitational Radiation from Primordial Turbulence}",
    eprint = "0705.1733",
    archivePrefix = "arXiv",
    primaryClass = "astro-ph",
    doi = "10.1103/PhysRevD.76.083002",
    journal = "Phys. Rev. D",
    volume = "76",
    pages = "083002",
    year = "2007"
}

@article{Turbulence3,
    author = "Dolgov, Alexander D. and Grasso, Dario and Nicolis, Alberto",
    title = "{Relic backgrounds of gravitational waves from cosmic turbulence}",
    eprint = "astro-ph/0206461",
    archivePrefix = "arXiv",
    doi = "10.1103/PhysRevD.66.103505",
    journal = "Phys. Rev. D",
    volume = "66",
    pages = "103505",
    year = "2002"
}

@article{Dvali:2022rgx,
    author = "Dvali, Gia and Valbuena-Berm\'udez, Juan Sebasti\'an",
    title = "{Erasure of strings and vortices}",
    eprint = "2212.07535",
    archivePrefix = "arXiv",
    primaryClass = "hep-th",
    doi = "10.1103/PhysRevD.107.035001",
    journal = "Phys. Rev. D",
    volume = "107",
    number = "3",
    pages = "035001",
    year = "2023"
}

@article{SolCoh1,
    author = "Dvali, Gia and Gomez, Cesar and Gruending, Lukas and Rug, Tehseen",
    title = "{Towards a Quantum Theory of Solitons}",
    eprint = "1508.03074",
    archivePrefix = "arXiv",
    primaryClass = "hep-th",
    doi = "10.1016/j.nuclphysb.2015.10.017",
    journal = "Nucl. Phys. B",
    volume = "901",
    pages = "338--353",
    year = "2015"
}

@article{Coh1,
    author = "Zhang, Wei-Min",
    editor = "Mitra, Asoke N.",
    title = "{Coherent states in field theory}",
    eprint = "hep-th/9908117",
    archivePrefix = "arXiv",
    pages = "297--323",
    month = "8",
    year = "1999"
}

@article{Berezhiani:2024pub,
    author = "Berezhiani, Lasha and Dvali, Gia and Sakhelashvili, Otari",
    title = "{Coherent states in gauge theories: Topological defects and other classical configurations}",
    eprint = "2411.11657",
    archivePrefix = "arXiv",
    primaryClass = "hep-th",
    doi = "10.1103/PhysRevD.111.065018",
    journal = "Phys. Rev. D",
    volume = "111",
    number = "6",
    pages = "065018",
    year = "2025"
}

@article{CGC1,
    author = "Dvali, Gia and Venugopalan, Raju",
    title = "{Classicalization and unitarization of wee partons in QCD and gravity: The CGC-black hole correspondence}",
    eprint = "2106.11989",
    archivePrefix = "arXiv",
    primaryClass = "hep-th",
    doi = "10.1103/PhysRevD.105.056026",
    journal = "Phys. Rev. D",
    volume = "105",
    number = "5",
    pages = "056026",
    year = "2022"
}

@article{CGC2,
    author = "Gelis, Francois and Iancu, Edmond and Jalilian-Marian, Jamal and Venugopalan, Raju",
    title = "{The Color Glass Condensate}",
    eprint = "1002.0333",
    archivePrefix = "arXiv",
    primaryClass = "hep-ph",
    doi = "10.1146/annurev.nucl.010909.083629",
    journal = "Ann. Rev. Nucl. Part. Sci.",
    volume = "60",
    pages = "463--489",
    year = "2010"
}

@article{Dvali:2018ytn,
    author = "Dvali, Gia and Eisemann, Lukas and Michel, Marco and Zell, Sebastian",
    title = "{Universe's Primordial Quantum Memories}",
    eprint = "1812.08749",
    archivePrefix = "arXiv",
    primaryClass = "hep-th",
    reportNumber = "LMU-ASC 82/18; MPP-2018-302",
    doi = "10.1088/1475-7516/2019/03/010",
    journal = "JCAP",
    volume = "03",
    pages = "010",
    year = "2019"
}

@article{Dvali:2011aa,
    author = "Dvali, Gia and Gomez, Cesar",
    title = "{Black Hole's Quantum N-Portrait}",
    eprint = "1112.3359",
    archivePrefix = "arXiv",
    primaryClass = "hep-th",
    doi = "10.1002/prop.201300001",
    journal = "Fortsch. Phys.",
    volume = "61",
    pages = "742--767",
    year = "2013"
}

@article{Dvali:2013eja,
    author = "Dvali, Gia and Gomez, Cesar",
    title = "{Quantum Compositeness of Gravity: Black Holes, AdS and Inflation}",
    eprint = "1312.4795",
    archivePrefix = "arXiv",
    primaryClass = "hep-th",
    doi = "10.1088/1475-7516/2014/01/023",
    journal = "JCAP",
    volume = "01",
    pages = "023",
    year = "2014"
}

@article{Dvali:2014gua,
    author = "Dvali, Gia and Gomez, Cesar",
    title = "{Quantum Exclusion of Positive Cosmological Constant?}",
    eprint = "1412.8077",
    archivePrefix = "arXiv",
    primaryClass = "hep-th",
    doi = "10.1002/andp.201500216",
    journal = "Annalen Phys.",
    volume = "528",
    pages = "68--73",
    year = "2016"
}

@article{Dvali:2017eba,
    author = "Dvali, Gia and Gomez, Cesar and Zell, Sebastian",
    title = "{Quantum Break-Time of de Sitter}",
    eprint = "1701.08776",
    archivePrefix = "arXiv",
    primaryClass = "hep-th",
    reportNumber = "LMU-ASC-08-17, MPP-2017-10",
    doi = "10.1088/1475-7516/2017/06/028",
    journal = "JCAP",
    volume = "06",
    pages = "028",
    year = "2017"
}

@article{Dvali:2011wk,
    author = "Dvali, Gia",
    title = "{Safety of Minkowski Vacuum}",
    eprint = "1107.0956",
    archivePrefix = "arXiv",
    primaryClass = "hep-th",
    month = "7",
    year = "2011"
}

@article{Dvali:2020etd,
    author = "Dvali, Gia",
    title = "{$S$-Matrix and Anomaly of de Sitter}",
    eprint = "2012.02133",
    archivePrefix = "arXiv",
    primaryClass = "hep-th",
    doi = "10.3390/sym13010003",
    journal = "Symmetry",
    volume = "13",
    number = "1",
    pages = "3",
    year = "2020"
}

@article{Bennett:1993pj,
    author = "Bennett, D. L. and Nielsen, Holger Bech",
    title = "{Predictions for nonAbelian fine structure constants from multicriticality}",
    eprint = "hep-ph/9311321",
    archivePrefix = "arXiv",
    reportNumber = "NBI-HE-93-22",
    doi = "10.1142/S0217751X94002090",
    journal = "Int. J. Mod. Phys. A",
    volume = "9",
    pages = "5155--5200",
    year = "1994"
}

@article{Coleman:1980aw,
    author = "Coleman, Sidney R. and De Luccia, Frank",
    title = "{Gravitational Effects on and of Vacuum Decay}",
    reportNumber = "SLAC-PUB-2463",
    doi = "10.1103/PhysRevD.21.3305",
    journal = "Phys. Rev. D",
    volume = "21",
    pages = "3305",
    year = "1980"
}

@article{Linde,
    author = "Linde, Andrei D.",
    title = "{Decay of the False Vacuum at Finite Temperature}",
    reportNumber = "LEBEDEV-81-265",
    doi = "10.1016/0550-3213(83)90072-X",
    journal = "Nucl. Phys. B",
    volume = "216",
    pages = "421",
    year = "1983",
    note = "[Erratum: Nucl.Phys.B 223, 544 (1983)]"
}

@article{Kobzarev,
    author = "Kobzarev, I. Yu. and Okun, L. B. and Voloshin, M. B.",
    title = "{Bubbles in Metastable Vacuum}",
    reportNumber = "ITEP-81-1974",
    journal = "Yad. Fiz.",
    volume = "20",
    pages = "1229--1234",
    year = "1974"
}

@article{Coleman:1977py,
    author = "Coleman, Sidney R.",
    title = "{The Fate of the False Vacuum. 1. Semiclassical Theory}",
    reportNumber = "HUTP-77-A004",
    doi = "10.1103/PhysRevD.16.1248",
    journal = "Phys. Rev. D",
    volume = "15",
    pages = "2929--2936",
    year = "1977",
    note = "[Erratum: Phys.Rev.D 16, 1248 (1977)]"
}

@article{MacroQ,
    author = "Dvali, Gia and Gomez, Cesar",
    title = "{Black Hole Macro-Quantumness}",
    eprint = "1212.0765",
    archivePrefix = "arXiv",
    primaryClass = "hep-th",
    month = "12",
    year = "2012"
}

@article{Dvali:2007hz,
    author = "Dvali, Gia",
    title = "{Black Holes and Large N Species Solution to the Hierarchy Problem}",
    eprint = "0706.2050",
    archivePrefix = "arXiv",
    primaryClass = "hep-th",
    doi = "10.1002/prop.201000009",
    journal = "Fortsch. Phys.",
    volume = "58",
    pages = "528--536",
    year = "2010"
}

@article{Dvali:2007wp,
    author = "Dvali, Gia and Redi, Michele",
    title = "{Black Hole Bound on the Number of Species and Quantum Gravity at LHC}",
    eprint = "0710.4344",
    archivePrefix = "arXiv",
    primaryClass = "hep-th",
    doi = "10.1103/PhysRevD.77.045027",
    journal = "Phys. Rev. D",
    volume = "77",
    pages = "045027",
    year = "2008"
}

@article{Guada:2020xnz,
    author = "Guada, Victor and Nemev{\v{s}}ek, Miha and Pintar, Matev{\v{z}}",
    title = "{FindBounce: Package for multi-field bounce actions}",
    eprint = "2002.00881",
    archivePrefix = "arXiv",
    primaryClass = "hep-ph",
    doi = "10.1016/j.cpc.2020.107480",
    journal = "Comput. Phys. Commun.",
    volume = "256",
    pages = "107480",
    year = "2020"
}

@article{Guada:2018jek,
    author = "Guada, Victor and Maiezza, Alessio and Nemev{\v{s}}ek, Miha",
    title = "{Multifield Polygonal Bounces}",
    eprint = "1803.02227",
    archivePrefix = "arXiv",
    primaryClass = "hep-th",
    doi = "10.1103/PhysRevD.99.056020",
    journal = "Phys. Rev. D",
    volume = "99",
    number = "5",
    pages = "056020",
    year = "2019"
}

@article{Kosowsky:1991ua,
    author = "Kosowsky, Arthur and Turner, Michael S. and Watkins, Richard",
    title = "{Gravitational radiation from colliding vacuum bubbles}",
    reportNumber = "FERMILAB-PUB-91-323-A",
    doi = "10.1103/PhysRevD.45.4514",
    journal = "Phys. Rev. D",
    volume = "45",
    pages = "4514--4535",
    year = "1992"
}

@article{Kosowsky:1992rz,
    author = "Kosowsky, Arthur and Turner, Michael S. and Watkins, Richard",
    title = "{Gravitational waves from first order cosmological phase transitions}",
    reportNumber = "FERMILAB-PUB-91-333-A-REV, FERMILAB-PUB-91-333-A",
    doi = "10.1103/PhysRevLett.69.2026",
    journal = "Phys. Rev. Lett.",
    volume = "69",
    pages = "2026--2029",
    year = "1992"
}

@article{Dvali:2021tez,
    author = "Dvali, Gia and Kaikov, Oleg and Berm{\'u}dez, Juan Sebasti{\'a}n Valbuena",
    title = "{How special are black holes? Correspondence with objects saturating unitarity bounds in generic theories}",
    eprint = "2112.00551",
    archivePrefix = "arXiv",
    primaryClass = "hep-th",
    doi = "10.1103/PhysRevD.105.056013",
    journal = "Phys. Rev. D",
    volume = "105",
    number = "5",
    pages = "056013",
    year = "2022"
}

@article{Dvali:2024hsb,
    author = "Dvali, Gia and Valbuena-Berm{\'u}dez, Juan Sebasti{\'a}n and Zantedeschi, Michael",
    title = "{Memory burden effect in black holes and solitons: Implications for PBH}",
    eprint = "2405.13117",
    archivePrefix = "arXiv",
    primaryClass = "hep-th",
    doi = "10.1103/PhysRevD.110.056029",
    journal = "Phys. Rev. D",
    volume = "110",
    number = "5",
    pages = "056029",
    year = "2024"
}

@article{Dvali:2018xpy,
    author = "Dvali, Gia",
    title = "{A Microscopic Model of Holography: Survival by the Burden of Memory}",
    eprint = "1810.02336",
    archivePrefix = "arXiv",
    primaryClass = "hep-th",
    month = "10",
    year = "2018"
}

@article{Dvali:2020wft,
    author = "Dvali, Gia and Eisemann, Lukas and Michel, Marco and Zell, Sebastian",
    title = "{Black hole metamorphosis and stabilization by memory burden}",
    eprint = "2006.00011",
    archivePrefix = "arXiv",
    primaryClass = "hep-th",
    doi = "10.1103/PhysRevD.102.103523",
    journal = "Phys. Rev. D",
    volume = "102",
    number = "10",
    pages = "103523",
    year = "2020"
}

@article{Dvali:2025sog,
    author = "Dvali, Gia",
    title = "{Swift Memory Burden in Merging Black Holes: how information load affects black hole's classical dynamics}",
    eprint = "2509.22540",
    archivePrefix = "arXiv",
    primaryClass = "hep-th",
    month = "9",
    year = "2025"
}

@article{Vachaspati:1984yi,
    author = "Vachaspati, Tanmay and Everett, Allen E. and Vilenkin, Alexander",
    title = "{Radiation From Vacuum Strings and Domain Walls}",
    reportNumber = "Print-84-0557 (TUFTS)",
    doi = "10.1103/PhysRevD.30.2046",
    journal = "Phys. Rev. D",
    volume = "30",
    pages = "2046",
    year = "1984"
}

@article{Berezin:1982ur,
    author = "Berezin, V. A. and Kuzmin, V. A. and Tkachev, I. I.",
    title = "{THIN WALL VACUUM DOMAINS EVOLUTION}",
    reportNumber = "IYaI-P-0258",
    doi = "10.1016/0370-2693(83)90630-5",
    journal = "Phys. Lett. B",
    volume = "120",
    pages = "91--96",
    year = "1983"
}

@article{Dvali:2012en,
    author = "Dvali, Gia and Gomez, Cesar",
    title = "{Black Holes as Critical Point of Quantum Phase Transition}",
    eprint = "1207.4059",
    archivePrefix = "arXiv",
    primaryClass = "hep-th",
    doi = "10.1140/epjc/s10052-014-2752-3",
    journal = "Eur. Phys. J. C",
    volume = "74",
    pages = "2752",
    year = "2014"
}

@article{Contri:2025eod,
    author = "Contri, Giacomo and Dvali, Gia and Sakhelashvili, Otari",
    title = "{Similarities in the evaporation of saturated solitons and black holes}",
    eprint = "2509.08049",
    archivePrefix = "arXiv",
    primaryClass = "hep-th",
    month = "9",
    year = "2025"
}

@article{Dvali:2013vxa,
    author = "Dvali, Gia and Flassig, Daniel and Gomez, Cesar and Pritzel, Alexander and Wintergerst, Nico",
    title = "{Scrambling in the Black Hole Portrait}",
    eprint = "1307.3458",
    archivePrefix = "arXiv",
    primaryClass = "hep-th",
    reportNumber = "LMU-ASC-50-13, LMU-ASC 50/13",
    doi = "10.1103/PhysRevD.88.124041",
    journal = "Phys. Rev. D",
    volume = "88",
    number = "12",
    pages = "124041",
    year = "2013"
}

@article{Yamamoto:1994te,
    author = "Yamamoto, Kazuhiro and Tanaka, Takahiro and Sasaki, Misao",
    title = "{Particle spectrum created through bubble nucleation and quantum field theory in the Milne Universe}",
    eprint = "gr-qc/9412011",
    archivePrefix = "arXiv",
    reportNumber = "KUNS-1305",
    doi = "10.1103/PhysRevD.51.2968",
    journal = "Phys. Rev. D",
    volume = "51",
    pages = "2968--2978",
    year = "1995"
}

@article{Aoyama:1982gg,
    author = "Aoyama, Hideaki",
    title = "{EFFECTS OF PAIR CREATION ON THE EXPANSION OF VACUUM BUBBLES}",
    reportNumber = "SLAC-PUB-3007, CALT-68-925-REV",
    doi = "10.1016/0550-3213(83)90590-4",
    journal = "Nucl. Phys. B",
    volume = "221",
    pages = "473--494",
    year = "1983"
}

@article{Rubakov:1984pa,
    author = "Rubakov, V. A.",
    title = "{PARTICLE CREATION DURING VACUUM DECAY}",
    reportNumber = "IYaI-P-0340",
    doi = "10.1016/0550-3213(84)90443-7",
    journal = "Nucl. Phys. B",
    volume = "245",
    pages = "481--516",
    year = "1984"
}

@article{Zeldovich:1974py,
    author = "Zeldovich, Ya. B.",
    title = "{Spontaneous processing in vacuum}",
    doi = "10.1016/0370-2693(74)90057-4",
    journal = "Phys. Lett. B",
    volume = "52",
    pages = "341--343",
    year = "1974"
}

@article{Bennett:1996vy,
    author = "Bennett, D. L. and Nielsen, Holger Bech",
    title = "{Gauge couplings calculated from multiple point criticality yield alpha**(-1) = 136.8 +- 9: At last the elusive case of U(1)}",
    eprint = "hep-ph/9607278",
    archivePrefix = "arXiv",
    reportNumber = "NBI-HE-96-29",
    doi = "10.1142/S0217751X9900155X",
    journal = "Int. J. Mod. Phys. A",
    volume = "14",
    pages = "3313--3385",
    year = "1999"
}

@article{Berezhiani:2020pbv,
    author = "Berezhiani, Lasha and Zantedeschi, Michael",
    title = "{Evolution of coherent states as quantum counterpart of classical dynamics}",
    eprint = "2011.11229",
    archivePrefix = "arXiv",
    primaryClass = "hep-th",
    doi = "10.1103/PhysRevD.104.085007",
    journal = "Phys. Rev. D",
    volume = "104",
    number = "8",
    pages = "085007",
    year = "2021"
}

@article{Berezhiani:2021gph,
    author = "Berezhiani, Lasha and Cintia, Giordano and Zantedeschi, Michael",
    title = "{Background-field method and initial-time singularity for coherent states}",
    eprint = "2108.13235",
    archivePrefix = "arXiv",
    primaryClass = "hep-th",
    doi = "10.1103/PhysRevD.105.045003",
    journal = "Phys. Rev. D",
    volume = "105",
    number = "4",
    pages = "045003",
    year = "2022"
}

@article{Berezhiani:2023uwt,
    author = "Berezhiani, Lasha and Cintia, Giordano and Zantedeschi, Michael",
    title = "{Perturbative construction of coherent states}",
    eprint = "2311.18650",
    archivePrefix = "arXiv",
    primaryClass = "hep-th",
    doi = "10.1103/PhysRevD.109.085018",
    journal = "Phys. Rev. D",
    volume = "109",
    number = "8",
    pages = "085018",
    year = "2024"
}

@article{Berezhiani:2025tkp,
    author = "Berezhiani, Lasha and Cintia, Giordano and Contri, Giacomo",
    title = "{Coherence and Quantum Stability of Relativistic Superfluid States}",
    eprint = "2509.21667",
    archivePrefix = "arXiv",
    primaryClass = "hep-th",
    month = "9",
    year = "2025"
}

@article{Berezhiani:2021zst,
    author = "Berezhiani, Lasha and Dvali, Gia and Sakhelashvili, Otari",
    title = "{de Sitter space as a BRST invariant coherent state of gravitons}",
    eprint = "2111.12022",
    archivePrefix = "arXiv",
    primaryClass = "hep-th",
    doi = "10.1103/PhysRevD.105.025022",
    journal = "Phys. Rev. D",
    volume = "105",
    number = "2",
    pages = "025022",
    year = "2022"
}

@article{Berezhiani:2024boz,
    author = "Berezhiani, Lasha and Dvali, Gia and Sakhelashvili, Otari",
    title = "{Consistent Canonical Quantization of Gravity: Recovery of Classical GR from BRST-invariant Coherent States}",
    eprint = "2409.18777",
    archivePrefix = "arXiv",
    primaryClass = "hep-th",
    month = "9",
    year = "2024"
}

@article{Dvali:2017nis,
    author = "Dvali, Gia",
    title = "{Area law microstate entropy from criticality and spherical symmetry}",
    eprint = "1712.02233",
    archivePrefix = "arXiv",
    primaryClass = "hep-th",
    doi = "10.1103/PhysRevD.97.105005",
    journal = "Phys. Rev. D",
    volume = "97",
    number = "10",
    pages = "105005",
    year = "2018"
}

@article{Dvali:2017ktv,
    author = "Dvali, Gia",
    title = "{Critically excited states with enhanced memory and pattern recognition capacities in quantum brain networks: Lesson from black holes}",
    eprint = "1711.09079",
    archivePrefix = "arXiv",
    primaryClass = "quant-ph",
    month = "11",
    year = "2017"
}

@article{Dvali:2018vvx,
    author = "Dvali, Gia",
    title = "{Black Holes as Brains: Neural Networks with Area Law Entropy}",
    eprint = "1801.03918",
    archivePrefix = "arXiv",
    primaryClass = "hep-th",
    doi = "10.1002/prop.201800007",
    journal = "Fortsch. Phys.",
    volume = "66",
    number = "4",
    pages = "1800007",
    year = "2018"
}

@article{Dvali:2018xoc,
    author = "Dvali, Gia",
    title = "{Classicalization Clearly: Quantum Transition into States of Maximal Memory Storage Capacity}",
    eprint = "1804.06154",
    archivePrefix = "arXiv",
    primaryClass = "hep-th",
    month = "4",
    year = "2018"
}

@article{Dvali:2018tqi,
    author = "Dvali, Gia and Michel, Marco and Zell, Sebastian",
    title = "{Finding Critical States of Enhanced Memory Capacity in Attractive Cold Bosons}",
    eprint = "1805.10292",
    archivePrefix = "arXiv",
    primaryClass = "quant-ph",
    reportNumber = "LMU-ASC 31/18; MPP-2018-112, LMU-ASC-31-18, MPP-2018-112",
    doi = "10.1140/epjqt/s40507-019-0071-1",
    journal = "EPJ Quant. Technol.",
    volume = "6",
    pages = "1",
    year = "2019"
}

@article{Ruffini:1971bza,
    author = "Ruffini, Remo and Wheeler, John A.",
    title = "{Introducing the black hole}",
    doi = "10.1063/1.3022513",
    journal = "Phys. Today",
    volume = "24",
    number = "1",
    pages = "30",
    year = "1971"
}

@article{Bekenstein:1972ny,
    author = "Bekenstein, J. D.",
    title = "{Transcendence of the law of baryon-number conservation in black hole physics}",
    doi = "10.1103/PhysRevLett.28.452",
    journal = "Phys. Rev. Lett.",
    volume = "28",
    pages = "452--455",
    year = "1972"
}

@article{Bekenstein:1971hc,
    author = "Bekenstein, Jacob D.",
    title = "{Nonexistence of baryon number for static black holes}",
    doi = "10.1103/PhysRevD.5.1239",
    journal = "Phys. Rev. D",
    volume = "5",
    pages = "1239--1246",
    year = "1972"
}

@article{Bekenstein:1972ky,
    author = "Bekenstein, J. D.",
    title = "{Nonexistence of baryon number for black holes. ii}",
    doi = "10.1103/PhysRevD.5.2403",
    journal = "Phys. Rev. D",
    volume = "5",
    pages = "2403--2412",
    year = "1972"
}

@article{Teitelboim:1972pk,
    author = "Teitelboim, C.",
    title = "{Nonmeasurability of the baryon number of a black-hole}",
    doi = "10.1007/BF02756471",
    journal = "Lett. Nuovo Cim.",
    volume = "3S2",
    pages = "326--328",
    year = "1972"
}

@article{Teitelboim:1972ps,
    author = "Teitelboim, C.",
    title = "{Nonmeasurability of the lepton number of a black hole}",
    doi = "10.1007/BF02826050",
    journal = "Lett. Nuovo Cim.",
    volume = "3S2",
    pages = "397--400",
    year = "1972"
}

@article{Hartle:1971qq,
    author = "Hartle, J. B.",
    title = "{Long-range neutrino forces exerted by kerr black holes}",
    doi = "10.1103/PhysRevD.3.2938",
    journal = "Phys. Rev. D",
    volume = "3",
    pages = "2938--2940",
    year = "1971"
}

@article{Kou:2025mvt,
    author = "Kou, Wei and Chen, Xurong",
    title = "{Probing Saturon-like Limits in QCD Systems}",
    eprint = "2511.14381",
    archivePrefix = "arXiv",
    primaryClass = "hep-ph",
    month = "11",
    year = "2025"
}

@article{Berezhiani:2016grw,
    author = "Berezhiani, Lasha",
    title = "{On Corpuscular Theory of Inflation}",
    eprint = "1610.08433",
    archivePrefix = "arXiv",
    primaryClass = "hep-th",
    doi = "10.1140/epjc/s10052-017-4672-5",
    journal = "Eur. Phys. J. C",
    volume = "77",
    number = "2",
    pages = "106",
    year = "2017"
}

@article{Dvali:2021jto,
    author = "Dvali, Gia",
    title = "{Bounds on quantum information storage and retrieval}",
    eprint = "2107.10616",
    archivePrefix = "arXiv",
    primaryClass = "hep-th",
    doi = "10.1098/rsta.2021.0071",
    journal = "Phil. Trans. A. Math. Phys. Eng. Sci.",
    volume = "380",
    number = "2216",
    pages = "20210071",
    year = "2021"
}

@article{Vilenkin:1984ib,
    author = "Vilenkin, Alexander",
    title = "{Cosmic Strings and Domain Walls}",
    reportNumber = "PRINT-84-0840 (TUFTS)",
    doi = "10.1016/0370-1573(85)90033-X",
    journal = "Phys. Rept.",
    volume = "121",
    pages = "263--315",
    year = "1985"
}

@article{vanDissel:2025xqn,
    author = "van Dissel, Fabio and Pujol{\`a}s, Oriol",
    title = "{Soft oscillons}",
    eprint = "2505.00890",
    archivePrefix = "arXiv",
    primaryClass = "hep-th",
    doi = "10.1007/JHEP08(2025)123",
    journal = "JHEP",
    volume = "08",
    pages = "123",
    year = "2025"
}

@article{GarciaTello:2024lie,
    author = "Garcia Tello, Pablo and Succi, Sauro",
    title = "{Cosmological implications of the minimum viscosity principle}",
    eprint = "2407.18960",
    archivePrefix = "arXiv",
    primaryClass = "hep-th",
    doi = "10.1142/S0129183124502504",
    journal = "Int. J. Mod. Phys. C",
    volume = "36",
    number = "07",
    pages = "2450250",
    year = "2025"
}

@article{ELLIS2017103,
    author = "Ellis, Joshua",
    title = "{TikZ-Feynman: Feynman diagrams with TikZ}",
    eprint = "1601.05437",
    archivePrefix = "arXiv",
    primaryClass = "hep-ph",
    doi = "10.1016/j.cpc.2016.08.019",
    journal = "Comput. Phys. Commun.",
    volume = "210",
    pages = "103--123",
    year = "2017"
}

@article{Vilenkin:1981bx,
    author = "Vilenkin, A.",
    title = "{Gravitational radiation from cosmic strings}",
    doi = "10.1016/0370-2693(81)91144-8",
    journal = "Phys. Lett. B",
    volume = "107",
    pages = "47--50",
    year = "1981"
}

@article{Martin:1996ea,
    author = "Martin, Xavier and Vilenkin, Alexander",
    title = "{Gravitational wave background from hybrid topological defects}",
    eprint = "astro-ph/9606022",
    archivePrefix = "arXiv",
    doi = "10.1103/PhysRevLett.77.2879",
    journal = "Phys. Rev. Lett.",
    volume = "77",
    pages = "2879--2882",
    year = "1996"
}

@article{Martin:1996cp,
    author = "Martin, Xavier and Vilenkin, Alexander",
    title = "{Gravitational radiation from monopoles connected by strings}",
    eprint = "gr-qc/9612008",
    archivePrefix = "arXiv",
    doi = "10.1103/PhysRevD.55.6054",
    journal = "Phys. Rev. D",
    volume = "55",
    pages = "6054--6060",
    year = "1997"
}

@article{Gleiser:1998na,
    author = "Gleiser, Marcelo and Roberts, Ronald",
    title = "{Gravitational waves from collapsing vacuum domains}",
    eprint = "astro-ph/9807260",
    archivePrefix = "arXiv",
    reportNumber = "DART-HEP-98-03",
    doi = "10.1103/PhysRevLett.81.5497",
    journal = "Phys. Rev. Lett.",
    volume = "81",
    pages = "5497--5500",
    year = "1998"
}

@article{Wachter:2024zly,
    author = "Wachter, Jeremy M. and Olum, Ken D. and Blanco-Pillado, Jose J.",
    title = "{More accurate gravitational wave backgrounds from cosmic strings}",
    eprint = "2411.16590",
    archivePrefix = "arXiv",
    primaryClass = "gr-qc",
    month = "11",
    year = "2024"
}

@article{Dvali:2022vwh,
    author = "Dvali, Gia and Valbuena-Berm{\'u}dez, Juan Sebasti{\'a}n and Zantedeschi, Michael",
    title = "{Dynamics of confined monopoles and similarities with confined quarks}",
    eprint = "2210.14947",
    archivePrefix = "arXiv",
    primaryClass = "hep-th",
    doi = "10.1103/PhysRevD.107.076003",
    journal = "Phys. Rev. D",
    volume = "107",
    number = "7",
    pages = "076003",
    year = "2023"
}

@article{Bogolyubsky:1976nx,
    author = "Bogolyubsky, I. L. and Makhankov, V. G.",
    title = "{On the Pulsed Soliton Lifetime in Two Classical Relativistic Theory Models}",
    reportNumber = "JINR-E2-9695",
    journal = "JETP Lett.",
    volume = "24",
    pages = "12",
    year = "1976"
}

@article{Bogolyubsky:1976sc,
    author = "Bogolyubsky, I. L. and Makhankov, V. G.",
    title = "{Dynamics of Heavy Spherically-Symmetric Pulsons}",
    reportNumber = "JINR-E2-10223",
    journal = "Pisma Zh. Eksp. Teor. Fiz.",
    volume = "25",
    pages = "120--123",
    year = "1977"
}

@article{Gleiser:1993pt,
    author = "Gleiser, Marcelo",
    title = "{Pseudostable bubbles}",
    eprint = "hep-ph/9308279",
    archivePrefix = "arXiv",
    reportNumber = "DART-HEP-93-05",
    doi = "10.1103/PhysRevD.49.2978",
    journal = "Phys. Rev. D",
    volume = "49",
    pages = "2978--2981",
    year = "1994"
}

@article{Kolb:1993hw,
    author = "Kolb, Edward W. and Tkachev, Igor I.",
    title = "{Nonlinear axion dynamics and formation of cosmological pseudosolitons}",
    eprint = "astro-ph/9311037",
    archivePrefix = "arXiv",
    reportNumber = "FERMILAB-PUB-93-335-A",
    doi = "10.1103/PhysRevD.49.5040",
    journal = "Phys. Rev. D",
    volume = "49",
    pages = "5040--5051",
    year = "1994"
}

@article{Copeland:1995fq,
    author = "Copeland, Edmund J. and Gleiser, M. and Muller, H. -R.",
    title = "{Oscillons: Resonant configurations during bubble collapse}",
    eprint = "hep-ph/9503217",
    archivePrefix = "arXiv",
    reportNumber = "SUSX-TH-95-3-3, FERMILAB-PUB-95-021-A, DART-HEP-95-01",
    doi = "10.1103/PhysRevD.52.1920",
    journal = "Phys. Rev. D",
    volume = "52",
    pages = "1920--1933",
    year = "1995"
}

@article{Michel:2023ydf,
    author = "Michel, Marco and Zell, Sebastian",
    title = "{The Timescales of Quantum Breaking}",
    eprint = "2306.09410",
    archivePrefix = "arXiv",
    primaryClass = "quant-ph",
    doi = "10.1002/prop.202300163",
    journal = "Fortsch. Phys.",
    volume = "71",
    pages = "2300163",
    year = "2023"
}

@article{Dvali:2017ruz,
    author = "Dvali, Gia and Zell, Sebastian",
    title = "{Classicality and Quantum Break-Time for Cosmic Axions}",
    eprint = "1710.00835",
    archivePrefix = "arXiv",
    primaryClass = "hep-ph",
    reportNumber = "LMU-ASC-59-17, LMU-ASC 59/17",
    doi = "10.1088/1475-7516/2018/07/064",
    journal = "JCAP",
    volume = "07",
    pages = "064",
    year = "2018"
}

\end{document}